\documentclass[letterpaper,twocolumn,twoside]{IEEEtran}
\usepackage{amsmath, amssymb, bm, cite, epsfig, psfrag}

\def\beq{\begin{equation}}
\def\eeq{\end{equation}}
\def\beqa{\begin{eqnarray}}
\def\eeqa{\end{eqnarray}}
\def\beqan{\begin{eqnarray*}}
\def\eeqan{\end{eqnarray*}}

\def\Z{{\mathbb{Z}}}

\def\R{{\mathbb{R}}}

\def\diag{\mathop{\mathrm{diag}}}
\def\x{\times}

\newtheorem{definition}{Definition}
\newtheorem{theorem}{Theorem}
\newtheorem{lemma}{Lemma}

\newtheorem{assumption}{Assumption}

\def\qhat{\widehat{q}}

\def\uhat{\widehat{u}}
\def\xhat{\widehat{x}}
\def\zhat{\widehat{z}}
\def\la{\leftarrow}
\def\ra{\rightarrow}

\def\SNR{\mbox{\small \sffamily SNR}}

\def\arr{\rightarrow}
\def\Exp{\mathbf{E}}
\def\var{\mbox{\bf var}}

\newcommand{\wbf}{\mathbf{w}}
\newcommand{\xbf}{\mathbf{x}}
\newcommand{\xbfhat}{\widehat{\mathbf{x}}}
\newcommand{\ybf}{\mathbf{y}}
\newcommand{\zbf}{\mathbf{z}}

\newcommand{\Abf}{\mathbf{A}}

\newcommand{\Sbf}{\mathbf{S}}

\newcommand{\Xbf}{\mathbf{X}}
\newcommand{\Ybf}{\mathbf{Y}}
\newcommand{\Zbf}{\mathbf{Z}}
\newcommand{\Zhat}{\widehat{Z}}

\newcommand{\muavg}{\mu}

\def\FIn{F_{\rm in}}
\def\MseIn{{\cal E}_{\rm in}}
\def\MseBarIn{\overline{\cal E}_{\rm in}}

\def\MseBarOut{\overline{\cal E}_{\rm out}}

\def\NIn{N_{\rm in}}
\def\NOut{N_{\rm out}}

\def\muInit{\mu_{\rm init}}
\def\muLo{\mu_{\rm lo}}
\def\muHi{\mu_{\rm hi}}

\def\xHatInit{\widehat{x}_{\rm init}}
\def\HGenie{H^{\rm genie}}

\def\negSpace{\hspace{-0.15in}}

\title{Estimation with Random Linear Mixing, Belief Propagation
and Compressed Sensing}
\author{Sundeep Rangan 
\thanks{S. Rangan (email: srangan@poly.com) is with
        Polytechnic Institute of New York University, Brooklyn, NY.}
}

\begin{document}


\setcounter{page}{1}

\maketitle
\begin{abstract}
We apply Guo and Wang's relaxed belief propagation
(BP) method to the estimation of a random vector from
linear measurements followed by
a componentwise probabilistic measurement channel.
Relaxed BP uses a Gaussian approximation in standard
BP to obtain significant computational savings
for dense measurement matrices.
The main contribution of this paper is to extend the
relaxed BP method and analysis
to general (non-AWGN) output channels.
Specifically, we present  detailed equations for implementing
relaxed BP for general channels and show that relaxed BP has an
identical asymptotic large sparse limit behavior as standard BP,
as predicted by the Guo and Wang's
state evolution (SE) equations.
Applications are presented to compressed sensing and estimation
with bounded noise.
\end{abstract}



\section{Introduction}
\begin{figure}
\begin{center}
  \includegraphics[width=3.5in]{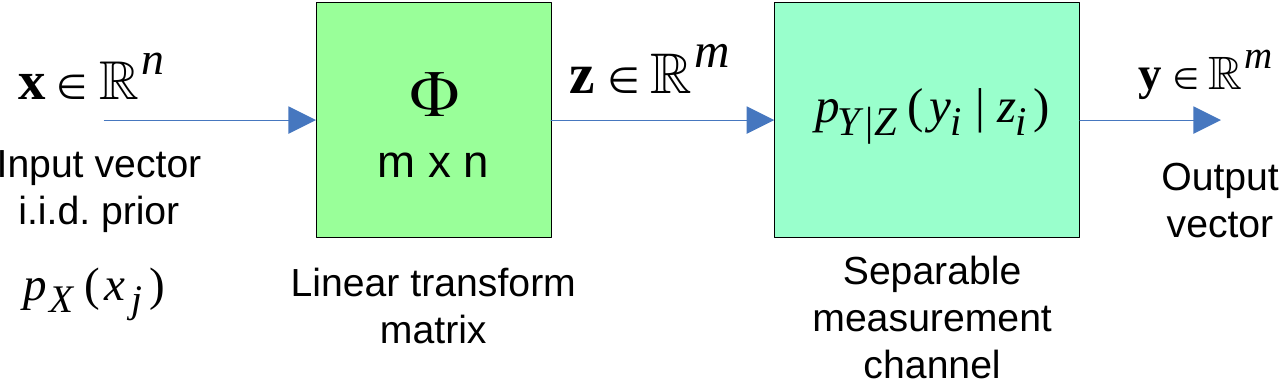}
\end{center}
\caption{Linear mixing estimation problem.  A random input vector $\xbf$
is transformed by a matrix $\Phi$, and the transformed vector $\zbf$
is passed through a separable measurement channel yielding a final output vector $\ybf$.
The linear mixing estimation problem is to estimate the input vector $\xbf$
from the output vector $\ybf$ given the transform matrix $\Phi$, prior $p_X(x_j)$
and measurement channel transition distribution $p_{Y|Z}(y_i|z_i)$. }
\label{fig:estModel}
\end{figure}

Consider the problem of estimating
a random vector $\xbf \in \R^n$ from the vector
$\ybf \in \R^m$ shown in Fig.\ \ref{fig:estModel}.
As depicted in the figure, the input vector $\xbf$ is first passed through
a linear transform,
\beq \label{eq:zPhix}
    \zbf = \Phi \xbf,
\eeq
where $\Phi \in \R^{m \x n}$ is a known transform matrix,
and then passed through an \emph{output channel} or \emph{measurement channel}
described by a conditional distribution $p_{\Ybf|\Zbf}(\ybf|\zbf)$.
Suppose that  the distributions of both the input vector $\xbf$
and output channel are \emph{separable} in that the probability
distributions factor as
\beq \label{eq:psep}
    p_{\Xbf}(\xbf) = \prod_{j=1}^n p_{X}(x_j), \ \ \
    p_{\Ybf|\Zbf}(\ybf|\zbf) =
        \prod_{i=1}^m p_{Y|Z}(y_i|z_i),
\eeq
where $p_X(x_j)$ and $p_{Y|Z}(y_i|z_i)$ are scalar distribution
functions, and $x_j$, $y_i$ and $z_i$ are the components
of the vectors $\xbf$, $\ybf$ and $\zbf$, respectively.

If $m=n$ and the mixing matrix $\Phi$ is the identity matrix,
then the problem of estimating the input vector $\xbf$
from the output vector $\ybf$ reduces
to $n$ scalar estimation problems.
However, for general $\Phi$, optimal estimation of
$\xbf$ is usually intractable because the transform matrix $\Phi$
``couples" or ``mixes" the $n$ components of $\xbf$
into the $m$ components of the output vector $\ybf$.
We thus call the problem of estimating the vector $\xbf$
from the coupled output vector $\ybf$
the \emph{linear mixing estimation problem}.

One natural approach to the linear mixing estimation problem is
\emph{belief propagation} (BP), which iteratively updates estimates
of the variables
based on message passing along a graph \cite{Pearl:88,WainwrightJ:08}.
In communications and signal processing,
BP is best known for its connections to
iterative decoding in turbo and LDPC codes \cite{McElieceMC:98,MacKay:99,MacKayN:97}.
However, while turbo and LDPC codes typically
involve computations over finite fields,
BP has also been successfully applied in a number of
problems with linear real-valued mixing, including
CDMA multiuser detection \cite{YoshidaT:06,GuoW:08},
lattice codes \cite{SommerFS:08} and
compressed sensing
\cite{BaronSB:10,GuoBS:09allerton,DonohoMM:09arxiv}.

A key theoretical justification for applying BP
to the specific problem of linear mixing estimation
came with the work of Montanari and Tse~\cite{MontanariT:06}.
That worked considered BP estimation of binary $\pm 1$ vectors
with AWGN measurements and
large sparse random mixing matrices.
In this setting, Montanari and Tse derived state evolution (SE) equations for
the mean-squared error of BP as a function
of the iteration number.  Their analysis revealed that
BP is asymptotically optimal in mean-square
when the SE equations have a unique fixed point.
The large sparse limit analysis was extended by Guo and Wang
first to general priors and power levels  \cite{GuoW:06},
and then to arbitrary (non-AWGN) output channels \cite{GuoW:07}.
These results provided the first rigorous conditions
for the optimality of BP for estimation with linear mixing
and confirmed earlier predictions given by the replica method from
statistical physics \cite{Tanaka:02,GuoV:05}.

Guo and Wang's work \cite{GuoW:06} also presented the
important result that the mean-square optimality of BP could
be achieved by a significantly simpler algorithm
that they called \emph{relaxed BP}.
One of the problems
of applying standard BP to the linear mixing estimation problem
is that the computations grow exponentially
with the density of the transform matrix $\Phi$.
Relaxed BP overcomes this problem by using a Gaussian approximation
of the messages to linearize the computations at the output nodes.
Gaussian approximations had been used in earlier BP-based methods
in CDMA multiuser detection
\cite{Kabashima:03,TanakaO:05,NeirottiS:05}
and also occasionally appear in the analysis and design of
LDPC codes \cite{ChungRU:01,Varshney:07}.

The main contribution of this paper is to extend
the relaxed BP method and analysis:
\begin{itemize}
\item \emph{Extensions to non-AWGN output channels:}
The relaxed BP algorithm described in Guo and Wang's first
paper \cite{GuoW:06} considers only AWGN output channels.
The second paper \cite{GuoW:07} considers arbitrary output
channels, but focuses on standard BP and only briefly mentions
how to apply the relaxed BP approximations.
In this paper, we work out the relaxed BP equations in full detail for
 general (non-AWGN) channels.
Moreover, we offer some additional simplifications (see Section \ref{sec:simpRBP})
to reduce the computations of relaxed BP even further.
The ability to incorporate non-AWGN channels
extends the scope of the relaxed BP method significantly.
For example, it enables the study of non-Gaussian noise
processes, as well
as discrete output channels that arise, for example, in pattern
classification problems.
We suggest some possible applications in Section \ref{sec:examples}.

\item \emph{Improved convergence analysis:}
A key result of Guo and Wang's state evolution (SE)
analysis in \cite{GuoW:06} and \cite{GuoW:07}
is that, when the measurement ratio $\beta = m/n$ is
sufficiently small, relaxed BP asymptotically achieves the minimum
mean-squared error (MSE) in the limit of large sparse
random mixing matrices.  Moreover, this minimum MSE is
described by a unique fixed point to the SE equations.

In this work, we extend the analysis to general $\beta$.
Specifically, we show that for \emph{any} $\beta$,
there are upper and lower
fixed point solutions to the SE equations.
The asymptotic
MSE of the relaxed BP method always converges
to the upper fixed point, while the lower fixed point
always provides a lower bound to the MSE of any estimator.
Hence, in the case that the fixed point
solution is unique, relaxed BP is asymptotically optimal.

\item \emph{Applications to compressed sensing and
bounded noise estimation:}
Although relaxed BP was originally developed for
CDMA multiuser detection, the method can be applied to
non-Gaussian estimation in a variety of applications.
In this paper, we simulate relaxed BP for compressed sensing
and estimation with bounded noise.
\end{itemize}

An algorithm closely related to relaxed BP is the recently
developed \emph{approximate message passing} (AMP) method
proposed in \cite{DonohoMM:09arxiv} and analyzed further in
\cite{BayatiM:10arxiv}.
The AMP algorithm generalizes the relaxed BP method and analysis
in the special case of AWGN measurements.
Unlike relaxed BP, the AMP algorithm can be applied with an
arbitrary scalar estimation function so that the prior on the
components of $\xbf$ do not need to be known.
Also, as we will discuss in Section \ref{sec:sparseLimAnal},
the analysis of relaxed BP is only valid under a certain
large sparse limit model.  This model is an approximation to the case
where $\Phi$ is dense.  The analysis of the
AMP algorithm in \cite{BayatiM:10arxiv}
provides rigorous results for dense measurement
matrices.  An interesting open problem is whether
the analysis of AMP can be extended to general output channels
considered here.

\subsection{Organization}
The remainder of this paper is organized as follows.
In Section \ref{sec:examples}, we introduce some specific examples
of the linear mixing estimation problem.
Section \ref{sec:bp} reviews how to apply standard BP to
estimation with linear mixing.
The relaxed BP algorithm is introduced in Section \ref{sec:rbp}.
The large sparse limit analysis is described in Section
\ref{sec:sparseLimAnal}.
Section \ref{sec:sim} presents some simple simulations of the algorithm
to validate the analytic results.
All the proofs are developed in appendices.

\section{Examples and Applications} \label{sec:examples}
The linear mixing model
is extremely general and can be applied in a range of circumstances.
We illustrate some simple examples for both the measurement
channel and prior on $\xbf$.

\subsection{Measurement Channel Examples} \label{sec:measChanEx}

\paragraph{AWGN output channel}  For an additive white Gaussian noise (AWGN)
output channel, the output vector $\ybf$ can be written as
\beq \label{eq:yzw}
    \ybf = \zbf + \wbf = \Phi\xbf + \wbf,
\eeq
where $\wbf$ is a zero mean, Gaussian i.i.d.\ random vector
independent of $\xbf$.
For this case, the corresponding channel transition probability
distribution is given by
\beq \label{eq:pyzAWGN}
    p_{Y|Z}(y_i|z_i) = \phi(y_i-z_i \, ; \, \mu_w),
\eeq
where $\mu_w > 0$ is the variance of the components of $\wbf$
and $\phi(v \, ; \, \mu)$ is the Gaussian distribution,
\beq \label{eq:Gaussian}
    \phi(v \, ; \, \mu) = \frac{1}{\sqrt{2\pi\mu}}
        \exp\left(-\frac{1}{2\mu}|v|^2\right).
\eeq
The AWGN channel
is precisely the model considered by Guo and Wang in their
original relaxed BP paper~\cite{GuoW:06}.

\paragraph{Non-Gaussian noise models}
Since the output channel can incorporate an arbitrary
separable distribution,
the linear mixing model can also include
the model (\ref{eq:yzw}) with non-Gaussian
noise vectors $\wbf$, provided the components of
$\wbf$ are i.i.d.
One interesting application for a non-Gaussian noise model
is to study the bounded noise that arises in quantization.
We will consider this application in the numerical simulations
in Section \ref{sec:bndSim}

\paragraph{Logistic channels}  A
quite different
channel is based on a \emph{logistic} output.
In this model, each output $y_i$ is $0$ or $1$,
where the probability
that $y_i=1$ is given by some sigmoidal function such as
\beq \label{eq:logitOut}
    p_{Y|Z}(y_i=1|z_i) = \frac{1}{1+a\exp(-z_i)},
\eeq
for some constant $a > 0$.  Thus, larger values of $z_i$
result in a higher probability that $y_i=1$.

This logistic model can be used in classification problems as follows \cite{Bishop:06}:
Suppose one is given $m$ samples, with each sample being labeled
as belonging to one of two classes.
Let $y_i = 0$ or 1 denote the class of sample $i$.
Also, suppose that the $i$th row of the transform matrix $\Phi$
contains a vector of $n$ data values associated with the $i$th sample.
Then, using a logistic channel model such as (\ref{eq:logitOut}),
the problem of estimating the vector $\xbf$ from the labels $\ybf$
and data $\Phi$ is equivalent to finding a linear
dependence on the data that classifies the samples between the
two classes.
This problem is often referred to as logistic regression
and the resulting vector $\xbf$ is called the regression vector.
By adjusting the prior on the components of $\xbf$, one
can then impose constraints on the components of $\xbf$
including, for example, sparsity constraints.

\subsection{Examples of Priors} \label{sec:priorEx}
\paragraph{Sparse priors and compressed sensing}
As discussed in the introduction, one class of priors
that we will consider in some detail in the simulations is sparse distributions.
A vector $\xbf$ is sparse if a large fraction of its components are zero
or close to zero.  Sparsity can be modeled probabilistically with a variety
of heavy-tailed distributions including Gaussian mixture models,
generalized Gaussians and Bernoulli distributions with a high probability of
the component being zero.
The estimation of sparse vectors with random linear measurements is
the basic subject of compressed sensing \cite{CandesRT:06-IT,Donoho:06,CandesT:06}
and fits naturally into the linear mixing framework.

\paragraph{Discrete distributions}
The linear mixing model can also incorporate discrete distributions on the
components of $\xbf$.
Discrete distribution arise often in communications problems
where discrete messages are modulated onto the components of $\xbf$.
The linear mixing with the transform matrix $\Phi$ comes into play
in CDMA spread spectrum systems and lattice codes mentioned above.

\section{Review of Standard Belief Propagation} \label{sec:bp}
\begin{figure}
\begin{center}
  \includegraphics[width=2in]{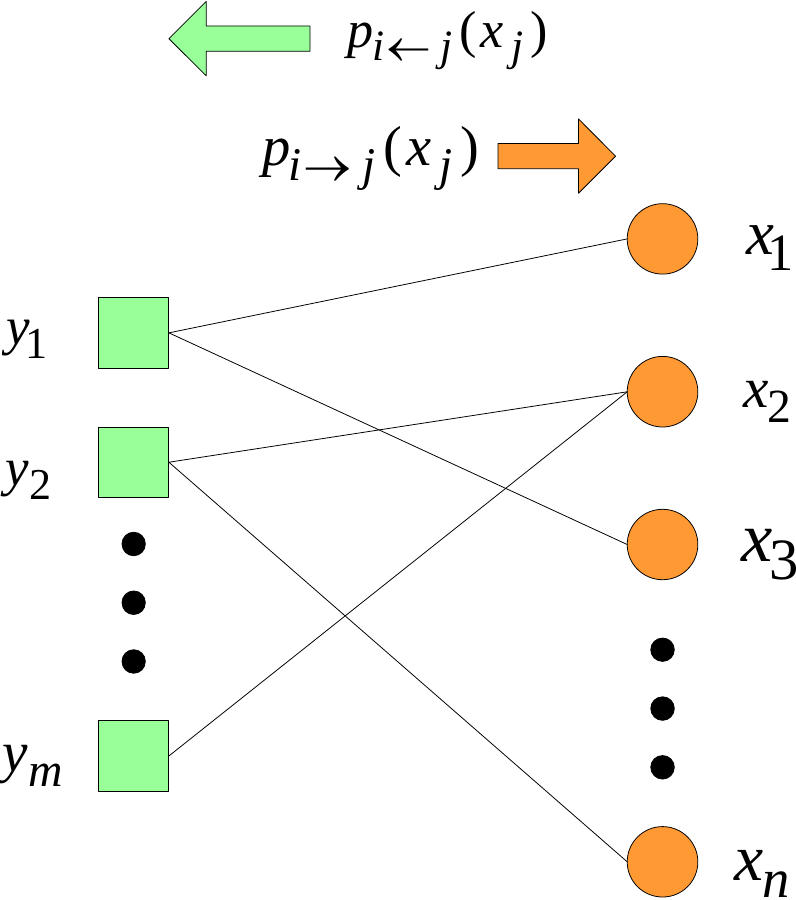}
\end{center}
\caption{Factor or Tanner graph for the linear mixing estimation problem. }
\label{fig:BPGraph}
\end{figure}

Before describing the relaxed BP algorithm,
it is useful to first review how standard BP would be applied
to the linear mixing estimation problem.
Standard BP associates with
the transform matrix $\Phi$ a bipartite graph $G = (V,E)$,
called the \emph{factor} or \emph{Tanner} graph
illustrated in Fig.~\ref{fig:BPGraph}.
The vertices $V$ in this graph consists of $n$ ``input" or
``variable" nodes
associated with the variables $x_j$, $j=1,\ldots,n$,
and $m$ ``output" or ``measurements" nodes
associated with the observations
$y_i$, $i=1,\ldots,m$.
There is an edge $(i,j) \in E$ between the input node $x_j$
and output node $y_i$ if and only if $\Phi_{i,j} \neq 0$.

Given this graph,
define the neighbor sets of the input and output nodes as
\begin{subequations} \label{eq:NsetDef}
\beqa
    \NIn(j) &=& \left\{ ~i~:~(i,j) \in E\right), \\
    \NOut(i) &=& \left\{ ~j~:~(i,j) \in E\right),
\eeqa
\end{subequations}
so that $\NIn(j)$ is the set of neighbors of the input index $j$,
and $\NOut(i)$ is the set of neighbors of the output index $i$.

The standard BP algorithm works by
iteratively passing ``messages" along the edges of this graph
represented as probability distributions on the variables $x_j$.
The messages are sometimes called \emph{beliefs}.
For the linear mixing estimation problem, the
standard BP algorithm can be described as follows:

\begin{enumerate}
\item \emph{Initialization:}  Set $t=1$ and initialize
the outgoing messages from the input nodes to
\beq \label{eq:pxBPinit}
    p^x_{i \la j}(t,x_j) = p^x_j(t,x_j) = p_X(x_j),
\eeq
for all input node indices $j$ and
edges $(i,j) \in E$.  This initialization simply
sets the messages to the priors on the variables $x_j$.

\item \emph{Mixing update:}  For each edge $(i,j) \in E$,
compute  $p^z_{i \ra j}(t,z_{i \ra j})$,
the distribution of the random variable
\beq \label{eq:zijBP}
    z_{i\ra j} = \sum_{r \in \NOut(i) \neq j} \negSpace \Phi_{ir}x_r,
\eeq
assuming the variables $x_r$ are independent with distributions
$x_r \sim p^x_{i \la r}(t,x_r)$.
Here, the sum is over indices $r \in \NOut(i)$ with $r \neq j$.
Also, for each $i$, compute $p^z_i(t,z_i)$, the distribution
of the random variable
\beq \label{eq:ziBP}
    z_i = \sum_{r \in \NOut(i)} \Phi_{ir}x_r.
\eeq

\item \emph{Output update:}  For each edge $(i,j) \in E$,
compute the likelihood function
\beqa
    p^u_{i \ra j}(t,u_i) & = & \int
        p_{Y|Z}(y_i\mid u_i + z_{i \ra j}) \nonumber \\
        & &\quad \, \times \, p^z_{i\ra j}(t,z_{i\ra j}) \, dz_{i\ra j}.\label{eq:puBP}
\eeqa
\item \emph{Input update:}  For each edge $(i,j) \in E$,
compute the distribution
\beq \label{eq:pxBP}
    p^x_{i \la j}(t+1,x_j) \propto
    p_X(x_j)\prod_{\ell \in \NIn(j) \neq i} \negSpace
        p^u_{\ell \ra j}(t,\Phi_{\ell j}x_j).
\eeq
Here, the $\propto$ sign indicates that the distribution is to be
normalized so that it has unit integral.
Also, compute the total distribution
\beq \label{eq:pxBP}
    p^x_{j}(t+1,x_j) \propto
    p_X(x_j)\prod_{\ell \in \NIn(j)} \negSpace
        p^u_{\ell \ra j}(t,\Phi_{\ell j}x_j).
\eeq
Increment $t=t+1$ and return to step 2
until a sufficient number of iterations have been performed.
\end{enumerate}

When the graph $G$ is
acyclic, then it can be shown that the distributions
$p^x_j(x_j)$ and $p^z_i(z_i)$
eventually converge to the true marginal distributions
of the random variables $x_j$ and $z_i$
given the observations $\ybf$.
However, for graphs with cycles, the BP algorithm in general
only returns an approximation to the true marginals.
An analysis of the BP algorithm is beyond the scope of this
work and is covered extensively elsewhere.  See, for example,
\cite{Pearl:88,WainwrightJ:08} and \cite{YedidiaFW:03}.

What is important here is to recognize the complexity
of the algorithm.  The difficult step is
the computations of the distributions of the variables $z_{i \ra j}$
and $z_i$ in (\ref{eq:zijBP}) and (\ref{eq:ziBP})
in the mixing update.
Suppose the output node $y_i$ has in-degree $d$.
That is, there are $d$ indices $j$ such that $\Phi_{ij}$ is
non-zero.  Then, the evaluation of the distributions on
$z_{i \ra j}$ and $z_i$ involve the integration over $d-1$
and $d$ components $x_r$.
Since the complexity of this computation grows exponentially
in $d$, the BP algorithm is only tractable when the
transform matrix $\Phi$ is sparse (i.e., $d$ is small).
The point of relaxed BP is to provide an approximation
to BP suitable for large, dense $\Phi$.

\section{Relaxed Belief Propagation} \label{sec:rbp}

\subsection{Scalar Estimation Functions} \label{sec:scaEst}

Before describing the relaxed BP algorithm, we need to define
certain functions related to scalar estimation problems at
the input and output nodes.  At the input nodes,
we consider the problem of estimating a scalar
random variable $x \sim p_X(x)$
from some scalar observation of the form
\beq \label{eq:qxv}
    q = x + v, \ \ \ v \sim {\cal N}(0,\mu),
\eeq
where $\mu > 0$ is a noise-level and $v$ is additive Gaussian
noise independent of $x$.
Let $\FIn(q, \mu)$ and $\MseIn(q,\mu)$ be the conditional
mean and variance of the random variable $x$
given the scalar observation $q$.
Although, the functions $\FIn(q,\mu)$ and $\MseIn(q,\mu)$
may not have closed form expressions, they can be evaluated with
one-dimensional integrals.

To analyze the output nodes, suppose that $z$ is
a scalar Gaussian random variable $z \sim {\cal N}(\zhat,\mu)$
and $y$ has the conditional distribution $p_{Y|Z}(y|z)$.
Let $p_{Y\mid \Zhat, \mu}(y | \zhat, \mu)$ be the likelihood
function
\beq \label{eq:pzy}
    p_{Y\mid \Zhat, \mu}(y | \zhat, \mu) =
    \int p_{Y\mid Z}(y|z)\phi(z - \zhat \, ; \, \mu) \, dz,
\eeq
where $\phi(z - \zhat \, ; \, \mu)$ is the Gaussian p.d.f.\
in (\ref{eq:Gaussian})
with mean $\zhat$ and variance $\mu$.
The relaxed BP algorithm is based on the
derivatives of log likelihood or \emph{score function}
\beq \label{eq:Drzdef}
    D_r(y,\zhat,\mu) = -\frac{\partial^r}{\partial \zhat^r}
        \log p_{y|\Zhat,\mu}(y|\zhat,\mu),
\eeq
for $r > 0$.
Again, this function can in general be evaluated numerically.

\subsection{Algorithm} \label{sec:rbpalgo}

We can now describe the relaxed BP algorithm.
The algorithm produces a sequence of estimates
$\xhat_j(t)$, $t=0,1,\ldots$ for each variable $x_j$
as well as estimates $\zhat_i(t)$ for each transformed variable $z_i$.
Several other intermediate estimates and variances
are also computed.  The steps are as follows:

\begin{enumerate}
\item \emph{Initialization:}  Set $t=1$, and
for every input node index $j$ and every $(i,j) \in E$, initialize
\begin{subequations} \label{eq:xmuInit}
\beqa
    \xhat_{i \la j}(t) &=& \xhat_j(t) \ = \ \xHatInit, \\
    \mu^x_{i \la j}(t) &=& \mu^x_j(t) \ = \ \muInit^x,
\eeqa
\end{subequations}
where $\xHatInit$ and $\muInit^x$ are the mean and variance of
the prior $p_X(x)$.
\item \emph{Output node, linear step:}  For every $(i,j) \in E$
compute
\begin{subequations} \label{eq:outLin}
\beqa
    \zhat_{i \ra j}(t) &=&
        \sum_{r \neq j}
        \Phi_{ir}\xhat_{i \la r}(t), \\
    \mu^z_{i \ra j}(t) &=&
        \sum_{r  \neq j}
        |\Phi_{ir}|^2\mu^x_{i \la r}(t). \label{eq:outLinMu}
\eeqa
\end{subequations}
Also compute $\zhat_i(t)$ and $\mu^z_i(t)$ similarly,
but with the summation over all $r \in \{1,\ldots,n\}$.

\item \emph{Output node, non-linear step:} For every $(i,j) \in E$
compute
\begin{subequations} \label{eq:outNL}
\beqa
    \uhat_{i \ra j}(t) &=& -\frac{
    D_1(y_i, \zhat_{i \ra j}(t), \mu^z_{i \ra j}(t))}
    {D_2(y_i, \zhat_{i \ra j}(t), \mu^z_{i \ra j}(t))}, \\
    \mu^u_{i \ra j}(t) &=& \frac{1}
    {D_2(y_i, \zhat_{i \ra j}(t),
    \mu^z_{i \ra j}(t))}, \label{eq:outNLMu}
\eeqa
\end{subequations}
where $D_r(y,\zhat,\mu)$ are the derivatives of the
negative log likelihood function in (\ref{eq:Drzdef}).

\item \emph{Input node, linear step:}  For every $(i,j) \in E$
compute
\begin{subequations} \label{eq:inLin}
\beqa
    \qhat_{i \la j}(t) &=& \mu^q_{i \la j}(t)
  \sum_{\ell \neq i}
    \frac{\Phi_{\ell j}^*
        \uhat_{\ell \ra j}(t)}{\mu^u_{\ell \ra j}(t)}, \hspace{0.2in}
        \label{eq:inLinq} \\
    \mu^q_{i \la j}(t) &=& \left(
    \sum_{\ell \neq i}
        \frac{|\Phi_{\ell j}|^2}{\mu^u_{\ell \ra j}(t)} \right)^{-1}. \label{eq:inLinMu}
\eeqa
\end{subequations}
Also, compute $\qhat_j(t)$ and $\mu^q_j(t)$ similarly,
but with the summation over all $\ell \in \{1,\ldots,m\}$.

\item \emph{Input node, non-linear step:}  For every $(i,j) \in E$
compute
\begin{subequations} \label{eq:inNL}
\beqa
    \xhat_{i \la j}(t+1) &=& \FIn(\qhat_{i \la j}(t),
        \mu^q_{i \la j}(t)), \\
    \mu^x_{i \la j}(t+1) &=& \MseIn(\qhat_{i \la j}(t),
        \mu^q_{i \la j}(t)). \label{eq:inNLMu}
\eeqa
\end{subequations}
Similarly, for every $j=1,\ldots,n$, compute $\xhat_j(t+1)$
and $\mu^x_j(t+1)$ using $\qhat_j(t)$ and $\mu^q_j(t)$.
Set $t = t+1$ and return to step 2.
\end{enumerate}

\subsection{Heuristic Justification} \label{sec:heuristic}

Although we will formally analyze the relaxed BP algorithm below,
it is useful to first provide a heuristic understanding
of the steps.
The relaxed BP algorithm is a simplification of the standard BP method
where only the means and variances of the probability distributions are passed.
Specifically, the terms $\xhat_{i \la j}(t)$
and $\mu^x_{i \la j}(t)$ are approximations of
the mean and variance of the distribution
$p^x_{i \la j}(t,x_j)$ in (\ref{eq:pxBP}) in the standard BP algorithm.
In step 1, these are initialized based on the prior $p_X(x_j)$.
The relaxed BP approximation does not assume that
$p^x_{i \la j}(t,x_j)$ itself is Gaussian.  However, relaxed BP does
assume that there is a sufficient number of
terms in the summation in (\ref{eq:zijBP}) that
$p^z_{i \ra j}(t,z_{i \ra j})$ of the standard BP algorithm
is well-approximated as Gaussian.  The terms
$\zhat_{i \ra j}(t)$ and $\mu^z_{i \ra j}(t)$ in (\ref{eq:outLin})
in step 2 of the relaxed BP algorithm
are the mean and variance of this Gaussian distribution.
Under this Gaussian assumption, the likelihood function
$p^u_{i \ra j}(t,u_i)$ in (\ref{eq:puBP}) is approximately given
by
\beq \label{eq:puapprox}
    p^u_{i \ra j}(t,u_i) \approx p_{Y|\Zhat,\mu}
    (y_i|\zhat_{i \ra j}(t),\mu^z_{i \ra j}(t)),
\eeq
where $p_{Y|\Zhat,\mu}(y|\zhat,\mu)$ is given in (\ref{eq:pzy}).
Then, using the derivatives in (\ref{eq:Drzdef}),
the second order approximation of (\ref{eq:puapprox})
is given by
\beqa
    \log p^u_{i \ra j}(t,u_i) & \approx &
    -\frac{1}{2\mu^u_{i \ra j}(t)}
    |u_i  - \uhat_{i \ra j}(t)|^2 \nonumber \\
    & &\quad + \, O(|u_i|^3) + \mbox{const}, \hspace{0.5in} \label{eq:puquad}
\eeqa
where $\uhat_{i \ra j}(t)$ and $\mu^u_{i \ra j}(t)$ are given in
(\ref{eq:outNL}) in step 3 of the relaxed BP algorithm
and the constant term does not depend on $u_i$.
Now summing the log likelihoods in (\ref{eq:puquad}),
the distribution $p^x_{i \la j}(t,x_j)$ in (\ref{eq:pxBP})
is given by
\beqa
    \lefteqn{ \log p^x_{i \la j}(t,x_j) = \mbox{const} + \log p_X(x_j) } \nonumber\\
    & & \qquad\qquad\qquad + \, \sum_{\ell \neq i} \log p^u_{\ell \ra j}(t,\Phi_{\ell j}x_j)
    \nonumber \\
        &\approx& \log p_X(x_j) -\frac{1}{2\mu^q_{i \la j}(t)}
        |x_j - \qhat_{i \la j}(t)|^2   + \mbox{const}, \hspace{0.2in}
        \label{eq:pxquad}
\eeqa
where $q_{i \la j}(t)$ and $\mu^q_{i \la j}(t)$ are the outputs
(\ref{eq:inLin}).  The approximation in (\ref{eq:pxquad})
is due to the fact that the sum of the $O(|\Phi_{ij}x_j|^3)$
terms is asymptotically negligible for large $n$.
This implies that
\[
    p^x_{i \la j}(t,x_j) \propto p_X(x_j)\phi(x_j -
        \qhat_{i \la j}(t) \, ; \, \mu^q_{i \la j}(t)),
\]
where again $\phi(\cdot\, ; \, \cdot)$ is the Gaussian distribution
in (\ref{eq:Gaussian}).  This implies that,
conditional on $x_j$,
$\qhat_{i \la j}(t)$ is distributed as ${\cal N}(x_j,\mu^q_{i \la j}(t))$.
The final step, step 5, uses the scalar estimation functions
in Section \ref{sec:scaEst} to compute
$\xhat_{i \la j}(t)$ and $\mu^x_{i \la j}(t)$, the mean
and variance of $x_j$ given $\qhat_{i \la j}(t)$.

\subsection{Algorithm Complexity and Simplifications} \label{sec:simpRBP}

We next consider the complexity of the relaxed BP algorithm.
The most computationally demanding steps are the
non-linear mean and variance computations in
$\FIn(\cdot)$, $\MseIn(\cdot)$ in (\ref{eq:inNL})
and the derivatives, $D_r(\cdot)$ of the log likelihood function in
(\ref{eq:outNL}).
Each of these functions can be computed by a one-dimensional numerical integral.
Moreover, each iteration of the relaxed BP algorithm requires exactly one evaluation
of the input node functions and one evaluation of the output log likelihood
derivatives per edge of the Tanner graph.
Thus, the computations grow linearly
with the density of the graph and unlike standard BP, the
relaxed BP algorithm is
tractable even for dense matrices $\Phi$.

The relaxed BP algorithm can
actually be further simplified with some small
approximations.
Following Tanaka and Okada's approximate BP algorithm in
\cite{TanakaO:05}, the relaxed BP
algorithm can be approximately implemented
using one evaluation of the input and output nonlinear functions per vertex,
as opposed to one evaluation per edge.

To implement this simplified relaxed BP algorithm,
we first assume that the outgoing variances are the same to all destinations.
Specifically, we replace the variance computations in the
relaxed BP algorithm with
\beqan
    \mu^z_{i \ra j}(t) &=& \mu^z_i(t) \\
    \mu^u_{i \ra j}(t) &=& \frac{1}{D_2(y_i,\zhat_i(t),\mu^z_i(t))} \\
    \mu^q_{i \la j}(t) &=& \mu^q_{j}(t) \\
    \mu^x_{i \la j}(t) &=& \mu^x_{j}(t),
\eeqan
where $\mu^q_j(t)$ and $\mu^x_j(t)$ are still computed as in the
relaxed BP algorithm.
In this way, the variance function $\MseIn(\cdot)$
and second derivative $D_2(y,\zhat,\mu)$
are each only computed once per vertex per iteration,
as opposed to once per edge.

To reduce the evaluations of the input node function $\FIn(\cdot)$,
we first observe from (\ref{eq:inLin}) and the definition
of $q_j(t)$ that
\[
    q_{i \la j}(t) = q_j(t) -
    \mu^q_{i \la j}(t)\frac{\Phi_{ij}^*u_{i \ra j}(t)}
        {\mu^u_{i \ra j}(t)}.
\]
Therefore, we can approximate the update in (\ref{eq:inNL}) with
\beqa
   \lefteqn{  \xhat_{i \la j}(t) = \FIn(\qhat_{i \la j}(t),\mu^q_{i \la j}(t)) }
    \nonumber \\
        &\approx & \FIn(\qhat_{i \la j}(t),\mu^q_{j}(t)) \nonumber \\
        &\approx & \FIn(\qhat_j(t),\mu^q_{j}(t)) \nonumber \\
        & & - \mu^q_{i \la j}(t)\frac{\Phi_{ij}^*u_{i \ra j}(t)}
        {\mu^u_{i \ra j}(t)}
        \left. \frac{\partial}{\partial q}\FIn(q,\mu^q_{j}(t))\right|_{q = \qhat_j(t)},
        \label{eq:xhatApprox}
\eeqa
where we have used a first order approximation for $\FIn(\cdot)$.
Moreover, the partial derivative can be evaluated with the following lemma.
\begin{lemma} \label{lem:mseDeriv}
The input MSE function defined in
Section \ref{sec:scaEst} satisfies
\beq \label{eq:mseInDeriv}
    \frac{\partial}{\partial q}\FIn(q,\mu)  =\frac{1}{\mu}\MseIn(q,\mu).
\eeq
\end{lemma}
\begin{IEEEproof}
See Appendix \ref{sec:mseDerivPf}.
\end{IEEEproof}

For the output node, we can use the fact that
\[
    \zhat_{i \ra j}(t) = \zhat_i(t) - \Phi_{ij}\xhat_{i \la j}(t).
\]
Therefore, we can make the approximation
\beqa
   \uhat_{i \ra j}(t) & = & D_1(y_i,\zhat_{i \ra j}(t),\mu^z_{i \ra j}(t))
    \nonumber \\
        &\approx & D_1(y_i,\zhat_{i \ra j}(t),\mu^z_i(t)) \nonumber \\
        &\approx & D_1(y_i,\zhat_i(t),\mu^z_i(t)) \nonumber \\
        & & \, - \, \Phi_{ij}\xhat_{i \la j}(t)
        D_2(y_i,\zhat_i(t),\mu^z_i(t)).
        \label{eq:uhatApprox}
\eeqa
Using (\ref{eq:xhatApprox}) and (\ref{eq:uhatApprox}), we only need to evaluate
the nonlinear functions $\FIn(\cdot)$ and $D_1(\cdot)$ once per edge.
The analysis that we will present later does not apply to the relaxed BP
algorithm with these approximations.
Nevertheless, we will see in the simulations that the approximate
relaxed BP
algorithms behave closely to the exact implementation.

\section{Large Sparse Limit Analysis}
\label{sec:sparseLimAnal}

\subsection{Modeling Assumptions}
We analyze the relaxed BP algorithm in the large sparse limit
developed in \cite{GuoW:06,GuoW:07,MontanariT:06}.
The large sparse limit model considers a sequence of problems parameterized by
$n$ and $d$.  For each $n$ and $d$,
the transform matrix $\Phi = \Phi(n,d) \in \R^{m \x n}$ is of the form
\beq \label{eq:PhiFactor}
    \Phi = \frac{1}{\sqrt{d}}\Abf\Sbf^{1/2},
    \ \ \ \Sbf = \diag(s_1,\ldots,s_n),
\eeq
where $\Abf \in \R^{m \x n}$ and
$\Sbf \in \R^{n \x n}$ are two matrices, and
$m = m(n)$ is a deterministic function of $n$.
The number of measurements is assumed to grow linearly in $n$ in that
\beq \label{eq:betaDef}
    \lim_{n \arr \infty} \frac{n}{m(n)} = \beta
\eeq
for some $\beta \geq 0$.

The components $s_j$ in (\ref{eq:PhiFactor}) are called the \emph{scale factors}
and are assumed to be i.i.d.\ with some probability distribution
$p_S(s_j)$ that does not depend on $n$ or $d$.
We assume $s_j > 0$ almost surely.
The diagonal scale factor matrix $\Sbf$ is used
to scale the powers of the components of $\xbf$.
Specifically, multiplication by $\Sbf^{1/2}$
scales the variance of the $j$th component of $\xbf$ by
a factor $s_j$.
In this way, the scale factors can be used to capture
variations in the power of the components
of $\xbf$ that are known \emph{a priori} at the estimator.
Variations in the power of $\xbf$ that are not known
to the estimator should be captured in the distribution of $\xbf$.

The matrix $\Abf$ is deterministic, and we evaluate the performance
of the relaxed BP algorithm on
a deterministic sequence of input and output indices $i=i(n,d)$ and $j=j(n,d)$.
The sequence of matrices $\Abf$ and indices $i$ and $j$ are assumed to
satisfy the following conditions:

\begin{assumption} \label{as:model}
Let $\Abf = \Abf(n,d)$ be a sequence of deterministic matrices
in the factorization of $\Phi = \Phi(n,d)$ in (\ref{eq:PhiFactor}).
Let $i=i(n,d)$ and $j=j(n,d)$ be a deterministic sequence of indices.
Let $t > 0$ be some fixed
iteration number of the relaxed BP algorithm.
Assume that $\Abf$, $i$ and $j$ satisfy the following:
\begin{itemize}
\item[(a)]  For every $n$ and $d$, $(i,j)$ is an edge
in the Tanner graph $G$ associated with $\Phi$.
\item[(b)]
The \emph{computation subgraphs} $G_i(t)$ and $G_j(t)$
of the Tanner graph
taken a depth of $2t$ hops from the output node $i$ and
input node $j$ are trees.  Precise definitions of these
computation subgraphs are given in Appendix \ref{sec:localTree}.
\item[(c)]  All the nodes in the subgraphs $G_i(t)$ and $G_j(t)$
have degrees bounded above by $d$.
\item[(d)] For all output nodes $\ell$ in the subgraph
$G_i(t)$, we have the limits
\begin{subequations}
\beqa
    \lim_{d \arr \infty} \lim_{n \arr \infty}
        \frac{1}{d}\sum_{r \in \NOut(\ell)} |a_{\ell r}|^2 &=& \beta,
           \label{eq:Asqr} \\
    \lim_{d \arr \infty} \lim_{n \arr \infty}
        \frac{1}{d^{3/2}}\sum_{r \in \NOut(\ell)} |a_{\ell r}|^3 &=& 0.   \label{eq:Acuber}
\eeqa
For all input nodes $r$ in the subgraph
$G_j(t)$, we have the limits
\beqa
    \lim_{d \arr \infty} \lim_{n \arr \infty}
        \frac{1}{d}\sum_{\ell \in \NIn(r)} |a_{\ell r}|^2 &=& 1,   \label{eq:Asql} \\
    \lim_{d \arr \infty} \lim_{n \arr \infty}
        \frac{1}{d^{3/2}}\sum_{\ell \in \NIn(r)} |a_{\ell r}|^3 &=& 0.   \label{eq:Acubel}
\eeqa
\end{subequations}
\end{itemize}
\end{assumption}

As in \cite{GuoW:06,GuoW:07,MontanariT:06}, 
the key assumption is that the Tanner graph $G$
associated with the transform matrix $\Phi$ is locally tree-like around
the components of interest $i$ and $j$.  The assumption is common
in the study of BP algorithms as it makes the messages independent.
This local tree-like property is only possible with the graph being sparse.
This sparsity assumption is brought out explicitly by bounding
the input and output degrees of the Tanner graph.

Assumption \ref{as:model} uses a deterministic model for the $\Abf$
as opposed to the random matrix model with
i.i.d.\ components studied in \cite{GuoW:06,GuoW:07,MontanariT:06}.
The deterministic model simplifies some of the proofs.  In particular,
the input and output degrees are deterministically
bounded as opposed to be bounded on average -- which simplifies some
of the convergence arguments.

In the large sparse limit analysis, we first let $n \arr \infty$
with $m$ growing linearly with $n$ and keeping $d$ fixed.
This enables the local-tree like properties.  We then
let $d \arr \infty$, which will enable the use of a Central Limit Theorem
approximation.

This order of limits is critical.
Unfortunately, to analyze dense matrices, one would like an analysis
where $d$ can grow with $n$.  Indeed, if the matrix is
completely dense, we would like $d = m(n)$.
Unfortunately, the
large sparse limit analysis that we rely on here
requires that we consider the two limits separately;
it thus represents an approximation to the actual problem.
Nevertheless, we will see in simulations that large sparse limit
analysis appears to predict the behavior with dense matrices as
well.

More sophisticated analysis techniques developed recently
in \cite{BayatiM:10arxiv} enable the study of dense matrices without
the order of limits above.  One possible avenue of future research would be
to see if that analysis can be applied to the relaxed BP algorithm
with general (non-AWGN) output channels as well.

\subsection{Large Sparse Limit Convergence} \label{eq:sparseConv}
Under the large sparse limit model, define the random vectors
\begin{subequations} \label{eq:thetaij}
\beqa
    \theta_{i \la j}^x(n,d,t) &=& (x_j,s_j,\xhat_{i \la j}(t),
        \mu^x_{i\la j}(t))  \\
   \theta_{j}^x(n,d,t) &=& (x_j,s_j,\xhat_{j}(t), \mu^x_{j}(t))
   \label{eq:thetajx}\\
    \theta_{i \ra j}^z(n,d,t) &=& (z_{i\ra j},\zhat_{i \ra j}(t),
        \mu^z_{i\ra j}(t))  \\
    \theta_{i}^z(n,d,t) &=& (z_i,\zhat_i(t),  \mu^z_i(t)),
\eeqa
\end{subequations}
where the dependence on $n$ and $d$ on the right-hand side of the
equations is implicit. Our goal is to describe the
large sparse limit behavior of these random vectors.

A key result of \cite{GuoW:07} is that the large sparse
limit behavior of BP is described by a set of simple
\emph{state evolution} (SE) equations, which can be described
as follows:
Given $\MseIn(q,\mu)$ in Section \ref{sec:scaEst}, define
\begin{subequations} \label{eq:MseBarIn}
\beqa
    \MseBarIn(\mu,s) &=& \Exp \left[ \MseIn(q,\mu/s) | s\right] \\
    \MseBarIn(\mu) &=& \Exp \left[ s\MseIn(q,\mu/s) \right],
\eeqa
\end{subequations}
where the expectation is taken over the scalar random variables
$s \sim p_S(s)$ and $q$ given by (\ref{eq:qxv}) with $x \sim p_X(x)$.
We will call $\MseBarIn(\mu)$
the \emph{input node MSE function}.
In addition to the works \cite{GuoW:06,GuoW:07},
this function appeared in
Guo and Verd\'u's replica analysis of MSE estimation \cite{GuoV:05}
and related works \cite{RanganFG:09arXiv,KabashimaWT:09arXiv}.
Variants also appear in the analysis of the AMP algorithm
\cite{DonohoMM:09arxiv,BayatiM:10arxiv}.

At the output node, let
\beq \label{eq:muzInit}
    \muInit^z = \beta \Exp(s) \muInit^x,
\eeq
where $\muInit^x$ is variance of $x_j$ according to the prior $p_X(x_j)$,
and the expectation is over $s \sim p_S(s)$.
Then, for $\mu \leq \muInit^z$,
Guo and Wang \cite{GuoW:07} define the \emph{output node MSE function} as
\beq \label{eq:MseBarOut}
    \MseBarOut(\mu) = \frac{1}{\Exp \left[ D_2(y,\zhat,\mu) \right]},
\eeq
where $D_2(y,\zhat,\mu)$ is the derivative
(\ref{eq:Drzdef}) of the score function.
The expectation in (\ref{eq:MseBarOut}) is taken over
\beq \label{eq:zzhat}
    (z,\zhat) \sim {\cal N}(0,P_z(\mu)),
\eeq
where $P_z(\mu)$ is the covariance matrix
\beq\label{eq:Pzmu}
        P_z(\mu)= \left(\begin{array}{cc}
            \muInit^z & \muInit^z-\mu \\ \muInit^z-\mu & \muInit^z-\mu
            \end{array} \right),
\eeq
and the conditional distribution of $y$ given $z$ is
given by $p_{Y|Z}(y|z)$.

Now consider the recursion
\begin{subequations} \label{eq:muSE}
\beqa
    \muavg^q(t) &=& \MseBarOut(\muavg^z(t)), \label{eq:muSEq} \\
    \muavg^x(t+1,s) &=& \MseBarIn(\muavg^q(t),s), \label{eq:muSEx} \\
    \muavg^z(t+1) &=& \beta \MseBarIn(\muavg^q(t)),
\eeqa
\end{subequations}
defined for $t \geq 1$.
We can also write (\ref{eq:muSE}) with the single equation
\beq \label{eq:muSEOne}
    \muavg^z(t+1) = \beta \MseBarIn\left[ \MseBarOut(\muavg^z(t)) \right].
\eeq
In \cite{GuoW:07}, the equations (\ref{eq:muSE}) (or the single equation
version (\ref{eq:muSEOne})) are called the \emph{state evolution}
equations for BP as they describe the evolution of the error variances.

We consider two possible initial conditions for this recursion:
one low value and one high value.
The low sequence will be initialized with $\mu^z(1) =\muLo^z(1) = 0$,
and the high sequence will be initialized with
$\mu^z(1) = \muHi^z(1) = \muInit^z$
in (\ref{eq:muzInit}). We will use the subscripts
as in $\muLo^z(t)$ and $\muHi^z(t)$ to differentiate
between the two sequences.

Now, for $t \in \Z^+$, let $\theta^x(t)$ be the random vector
\beq \label{eq:thetax}
    \theta^x(t) = \left(x,s,\FIn(q,\mu), \MseIn(q,\mu) \right),
\eeq
where $x \sim p_X(x)$, $s \sim p_S(s)$, $q$ is distributed as (\ref{eq:qxv}),
and $\mu = \mu^q(t-1)/s$
with  $\mu^q(t-1)$ being the (deterministic) quantity in the state evolution (SE)
equation (\ref{eq:muSE}).
To initialize, let
\beq \label{eq:thetaxInit}
    \theta^x(1) = \left(x,s,\xHatInit, \muInit^x \right),
\eeq
where $\xHatInit$ and $\muInit^x$ are the mean
and variance of the prior of $p_X(x)$.
We will see below that when we use $\mu^q(t) = \muHi^q(t)$,
the SE output with the ``high" initial condition
$\theta^x(t)$ describes the limit of the random vector
$\theta_{i \la j}^x(n,d,t)$ in (\ref{eq:thetaij}).
When $\mu^q(t) = \muLo^q(t)$, $\theta^x(t)$ is the limit
of a related quantity for a certain ``genie-aided" algorithm
(see Appendix \ref{sec:genieAlgo}).

Also, for all $t \in \Z^+$, define the random vector
\beq \label{eq:thetaz}
    \theta^z(t) = (z,\zhat,\mu^z(t)),
\eeq
where $\mu^z(t)$ is the output of the state evolution equations (\ref{eq:muSE}) and
$(z,\zhat) \sim {\cal N}(0,P_z(\mu^z(t)))$.

\begin{theorem}  \label{thm:thetaLim}
Consider the relaxed BP algorithm under the large sparse limit model above
with transform matrix $\Phi$ and indices $i$ and $j$ satisfying
Assumption \ref{as:model} for some
\emph{fixed} iteration number $t$.  Then:
\begin{itemize}
\item[(a)] The random vectors in (\ref{eq:thetaij}) converge in distribution as follows:
\begin{subequations} \label{eq:thetaLim}
\beqa
    \lim_{d \arr \infty} \lim_{n \arr \infty}
        \theta_{i \la j}^x(n,d,t) &=& \theta^x(t) \label{eq:thetaxLim} \\
    \lim_{d \arr \infty} \lim_{n \arr \infty}
        \theta_{j}^x(n,d,t) &=& \theta^x(t) \label{eq:thetaxTotLim} \\
    \lim_{d \arr \infty} \lim_{n \arr \infty}
        \theta_{i \la j}^z(n,d,t) &=& \theta^z(t) \label{eq:thetazLim} \\
    \lim_{d \arr \infty} \lim_{n \arr \infty}
        \theta_{i}^z(n,d,t) &=& \theta^z(t), \label{eq:thetazTotLim}
\eeqa
\end{subequations}
where the random vectors $\theta^x(t)$ and $\theta^z(t)$ are
defined as above with $\mu^q(t) = \muHi^q(t)$ and
$\mu^z(t) = \muHi^z(t)$.
\item[(b)]  The error variances satisfy the limits
\begin{subequations} \label{eq:mseLim}
\beqa
    \lim_{d \arr \infty} \lim_{n \arr\infty}
        \Exp\left[ |x_j-\xhat_j(t)|^2 | s_j=s\right] =
         \muHi^x(t,s) , \label{eq:mseLimx} \\
    \lim_{d \arr \infty} \lim_{n \arr\infty}
        \Exp\left[ |z_i-\zhat_i(t)|^2\right]
        = \muHi^z(t), \hspace{0.5in}
        \label{eq:mseLimz}
\eeqa
\end{subequations}
where $\muHi^x(t,s)$ and $\muHi^z(t)$ are the output of the SE
equations (\ref{eq:muSE}) with the ``hi" initial condition.
\item[(c)] The minimum conditional error variance
of $x_j$ and $z_i$ given $\Phi$ and $\ybf$ satisfy the
asymptotic lower bounds
\begin{subequations} \label{eq:mseLimLo}
\beqa
    \lim_{d \arr \infty} \lim_{n \arr\infty}
        \Exp\left[ \var(x_j|\ybf,\Phi) | s_j=s\right] \geq
         \muLo^x(t,s) , \label{eq:mseLimxLo}\\
    \lim_{d \arr \infty} \lim_{n \arr\infty}
        \Exp\left[ \var(z_i|\ybf,\Phi)\right]
        \geq \muLo^z(t), \hspace{0.5in}
        \label{eq:mseLimzLo}
\eeqa
\end{subequations}
where $\muLo^x(t,s)$ and $\muLo^z(t)$ are the output of the SE
equations (\ref{eq:muSE}) with the ``lo" initial condition.
\end{itemize}
\end{theorem}
\begin{IEEEproof}
See Appendices \ref{sec:thetaLimAPf} and \ref{sec:thetaLimBCPf}.
\end{IEEEproof}

\medskip
The performance bounds in parts (a) and (b) are
largely identical to the results in
\cite{GuoW:07} except that they apply to relaxed BP instead
of BP\@. This is our main result:
in the large sparse limit model, relaxed BP
and standard BP have the identical asymptotic behavior.
The lower bound in part (c) of the theorem is also very close
to results in \cite{GuoW:07} and just repeated here for
completeness.

\begin{figure}
\begin{center}
  \includegraphics[height=2in,width=3.25in]{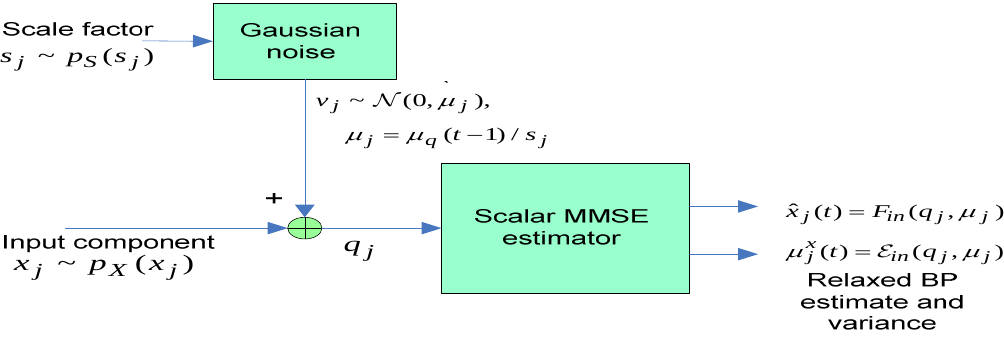}
\end{center}
\caption{Equivalent scalar model of the relaxed BP algorithm.
The asymptotic behavior of the
relaxed BP estimate $\xhat_j(t)$ of a component $x_j$ is identical
to the output of an MMSE estimator
with AWGN noise.  The effective noise is scaled by the factor $s_j$ corresponding to
the component $x_j$.}
\label{fig:scalarModel}
\end{figure}

Part (a) of the theorem
provides a simple scalar characterization for this asymptotic
behavior.
Specifically, using the definition of $\theta^x(t)$ in (\ref{eq:thetax}),
Theorem \ref{thm:thetaLim} shows that the componentwise behavior of
the relaxed BP follows a \emph{scalar equivalent model}
as shown in Fig.\ \ref{fig:scalarModel}:
The component $x_j$ is first corrupted by Gaussian noise
yielding a noisy component $q_j$.  The relaxed BP estimate $\xhat_j(t)$
then behaves identically to the optimal scalar MMSE estimate of
$x_j$ from the AWGN measurement $q_j$.
From this scalar equivalent joint distribution of the
components and their estimates,
one can compute any componentwise separable performance metric such
as mean-squared error or probability of detection.

The effective Gaussian noise levels in the scalar models are
described by $\muHi^z(t)$ and $\muHi^q(t)$
from the state evolution equations (\ref{eq:muSE}).
Since the state evolution equations can be evaluated easily with numerical
integration, Theorem~\ref{thm:thetaLim} thus
provides a simple, computationally-tractable
method for exactly characterizing the performance of the relaxed BP algorithm.

Part (b) shows that the SE outputs
$\muHi^x(t,s)$ and $\muHi^z(t)$ respectively describe
the asymptotic estimation error on the components $x_j$
and prediction error on the outputs $z_i$.
Part (c) provides corresponding lower bounds on these
error variances for any estimator.

\subsection{Convergence over Iteration and Mean-Square
Optimality} \label{eq:convIter}
Theorem \ref{thm:thetaLim} describes the asymptotic behavior of the
relaxed BP algorithm for a \emph{fixed} iteration number $t$.
Our second result describes the behavior of the relaxed BP estimates
as  $t \arr \infty$.

\begin{theorem} \label{thm:convIter}  Consider the state
evolution equations (\ref{eq:muSE}).
Suppose that $\MseBarIn(\mu)$ and $\MseBarOut(\mu)$
are continuous.   Then, the SE equations have at least
one fixed point with $0 \leq \mu^z \leq \muInit^z$.  Also:
\begin{itemize}
\item[(a)] With the ``hi" initial condition,
$\mu^z(1) = \muHi^z(1) = \muInit^z$, the sequences
$\muHi^z(t)$, $\muHi^x(t,s)$ and
$\muHi^q(t)$ decrease monotonically to the largest
fixed point of the SE equations (\ref{eq:muSE}).
\item[(b)] With the ``lo" initial condition,
$\mu^z(1) = \muLo^z(1) = 0$, the sequences
$\muLo^z(t)$, $\muLo^x(t,s)$ and
$\muLo^q(t)$ increase monotonically to the smallest
fixed point of the SE equations (\ref{eq:muSE}).
\end{itemize}
\end{theorem}
\begin{IEEEproof}
See Appendix \ref{sec:convIterPf}.
\end{IEEEproof}

The theorem is similar to the
convergence result in \cite{GuoW:06} except that it applies to all
$\beta$.
The importance of the
result is that it shows that relaxed BP provably converges
in the limit of large iterations, and the asymptotic
error variance of relaxed BP and the corresponding
error lower bounds are both fixed points of the SE equations.
A corollary of this result is that,
when the fixed points of the SE equations (\ref{eq:muSE})
are unique, the error variance of relaxed BP and the corresponding
lower bound agree.  The result thus gives an easily verifiable
condition under which relaxed BP is asymptotically
mean-square optimal.

\section{Numerical Simulations} \label{sec:sim}

The large sparse limit analysis
of the relaxed BP algorithm in Section \ref{sec:sparseLimAnal}
is theoretically exact only in the asymptotic limit of large dimensions.
Also, the analysis assumes a certain scaling where the measurement matrix $\Phi$
remains sparse.  To test the accuracy of the large sparse limit model for
finite problems with dense $\Phi$, we conducted the following simple numerical
experiments.

\subsection{Gauss-Bernoulli Prior with an AWGN Output Channel}
In the first experiment, the vector $\xbf$ was generated with
i.i.d.\ components $x_j$ with a Gauss-Bernoulli distribution given by
\beq \label{eq:gmixDist}
    x_j \sim \left\{ \begin{array}{ll}
        {\cal N}(0,1/\rho) & \mbox{with prob } = \rho, \\
        0 & \mbox{with prob } = 1-\rho. \\
    \end{array} \right.
\eeq
Here $\rho$ is the \emph{sparsity ratio}
and represents the average fraction of non-zero components
in $\xbf$.   The experiments below used the value $\rho = 0.1$.
We chose this Gaussian mixture model since it is
a simple example of a sparse prior used in compressed sensing.
This prior is also used in the numerical validation of the replica
method in \cite{RanganFG:09arXiv}.

The components of the measurement matrix $\Phi$ were generated as i.i.d.\
zero-mean Gaussians.  Even though this matrix is dense, we will see that the
large sparse limit analysis predicts the behavior of the relaxed BP estimator well.

For the measurement channel in this first experiment,
we assumed an AWGN output channel (\ref{eq:yzw}),
where the noise $\wbf$ also has i.i.d.\ zero-mean Gaussian components.
 The noise variance, $\mu_w$, of the components of $\wbf$
was selected such that $\SNR_0$ = 10 dB, where $\SNR_0$ is the signal-to-noise
ratio,
\beq \label{eq:SNRdef}
    \SNR_0 = 10 \log_{10}\left( \frac{\Exp\|\Phi\xbf\|^2}{n \mu_w}
    \right).
\eeq
As discussed in \cite{RanganFG:09arXiv}, $\SNR_0$ is the effective SNR
that an estimator would see in estimating any one component of $x_j$
with the other $n-1$ components of $\xbf$ known.

\begin{figure}
\begin{center}
  \includegraphics[width=3.5in]{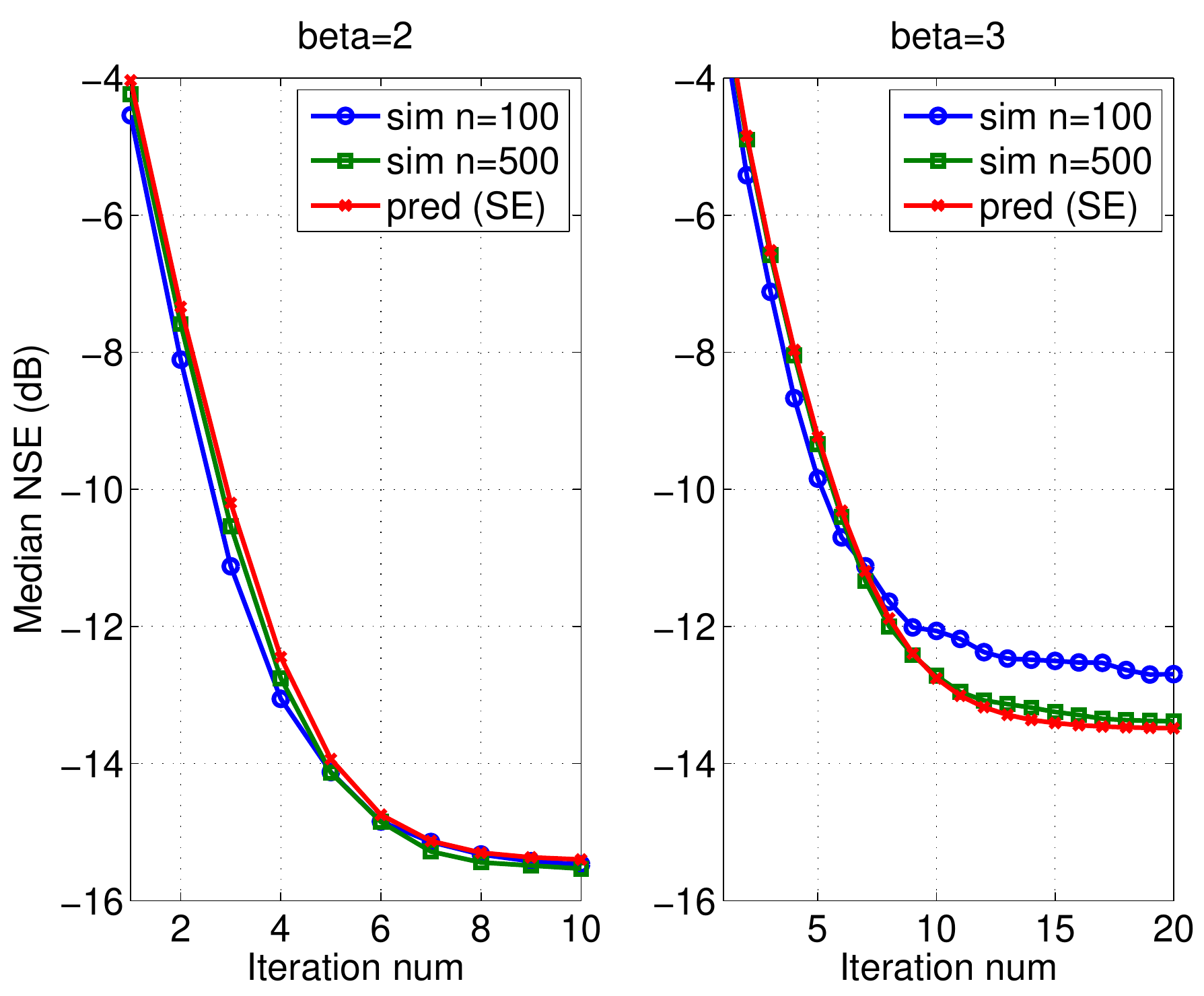}
\end{center}
\caption{State evolution prediction.
The normalized squared-error as predicted by the state evolution
equations (\ref{eq:muSE}) as a function of the iteration
number is compared against the simulated value for $n=100$ and $500$.
The simulation is based on a sparse Gauss-Bernoulli model.  See
text for details. }
\label{fig:gmixDE}
\end{figure}

Fig.\ \ref{fig:gmixDE} shows the median normalized squared-error (NSE)
as a function of the iteration number in the relaxed BP algorithm for this model.
In all the numerical experiments, we used the relaxed BP algorithm with the
simplifications described in Section \ref{sec:simpRBP}.
The simulation was conducted with 1000
random realizations of the problem, and for each
realization, we measured the NSE given by
\[
    \mbox{NSE} = 10 \log_{10}\left( \frac{
    \|\xbfhat - \xbf\|^2}{\Exp\|\xbf\|^2}
    \right),
\]
where $\xbfhat$ is the estimate of $\xbf$.  The NSE
represents the average error over the $n$ components of $\xbf$.
Fig.\ \ref{fig:gmixDE} plots the median of these NSE
values over the 1000 Monte Carlo trials.
The figure shows the median NSE for vector dimensions
$n=100$ and $500$ and $\beta = n/m = 2$ and $3$.

The points marked ``pred (SE)" are the NSE values as predicted
by the state evolution equations (\ref{eq:muSE}).
We see that the state evolution equations,
which are theoretically exact for infinite $n$,
provide an excellent match (within $0.1$ dB)
with the simulated values when $n=500$.
At the shorter length of $n=100$, the SE equations still
provide a good match, although there is a
small steady error of about 0.2 dB when $\beta = 2$ and 0.8 dB
when $\beta = 3$.

\begin{figure}
\begin{center}
  \includegraphics[width=3.5in]{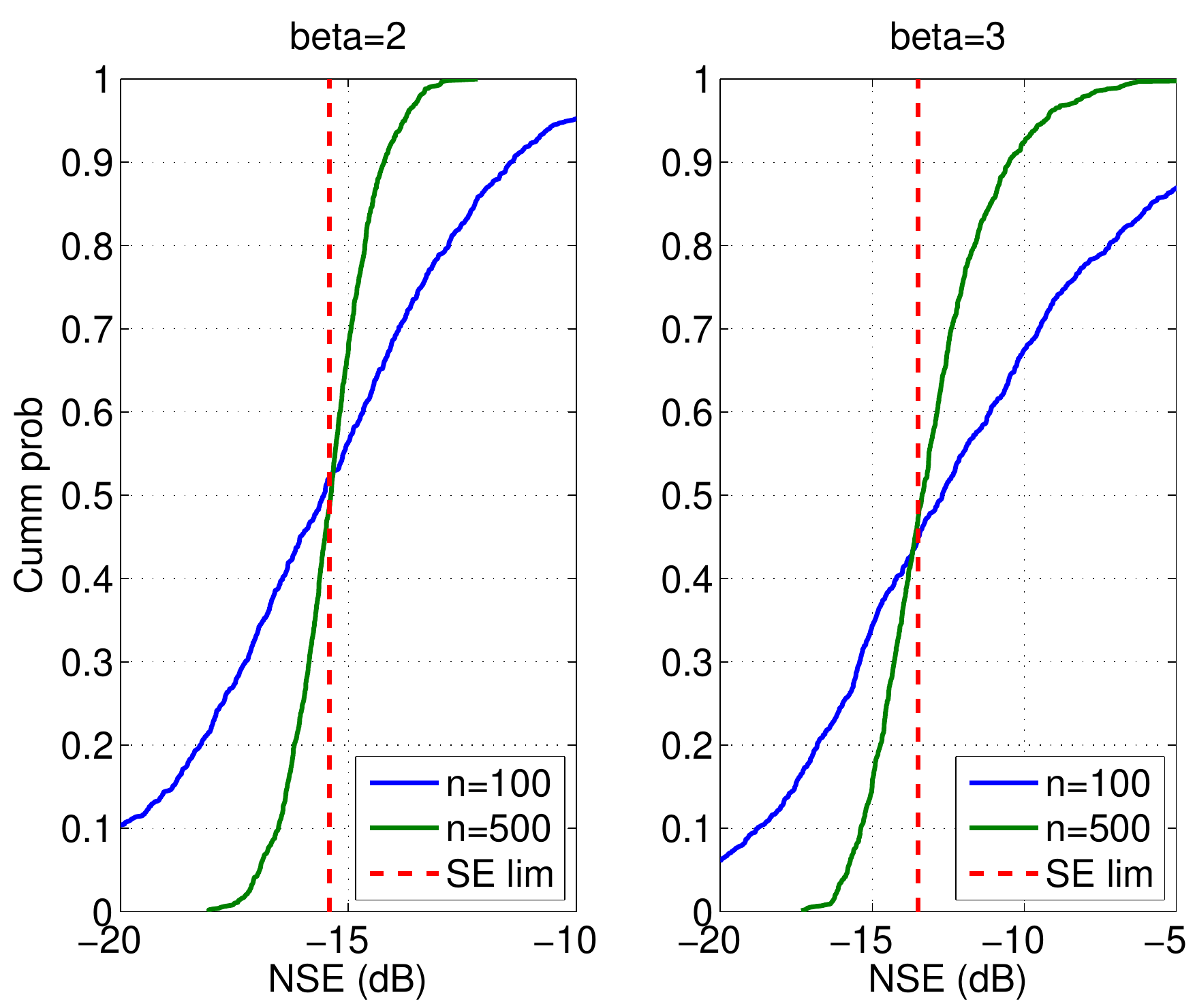}
\end{center}
\caption{Performance variation.  Plotted is the simulated CDF of the
NSE, averaged over the components in the vector $\xbf$,
but accounting for variations in the random problem parameters.
The CDF for the value of $n=500$ shows less variation than $n=100$
and closer to concentration around the state evolution limit (SE lim).
}
\label{fig:gmixMseCdf}
\end{figure}

In Fig.\ \ref{fig:gmixDE}, we plotted the median NSE
since there is actually considerable variation in the NSE
over the random realizations of the problem parameters.
To illustrate the degree of variability, Fig.\ \ref{fig:gmixMseCdf}
shows the CDF of the NSE values over the 1000 Monte Carlo trials.
We see that there is a large variation in the NSE, especially
at the smaller dimension $n=100$.
This means that although the median performance
may be good, there is still a significant chance that the
algorithm could perform well below the median on any
particular realization.

As one might expect, at the higher dimension of $n=500$,
the level of variability is reduced and performance begins to concentrate around
the density evolution limit.  However, even at $n=500$, the
variation is not insignificant.  As a result, caution should be exercised in
using the SE predictions with short to medium block lengths.

\begin{figure}
\begin{center}
  \includegraphics[width=3.5in]{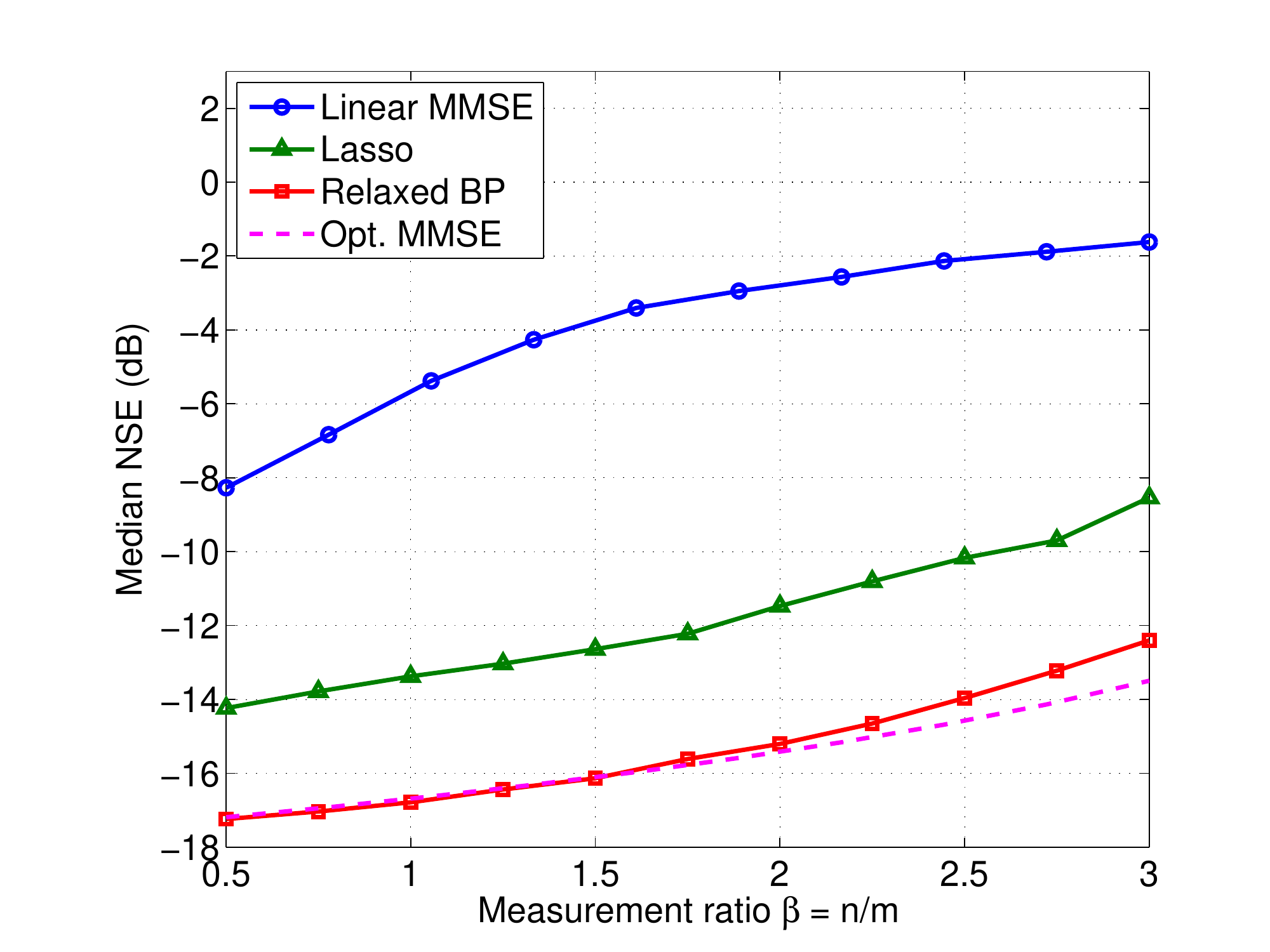}
\end{center}
\caption{Performance comparison of relaxed BP with other
sparse estimation methods.
}
\label{fig:gmixMethComp}
\end{figure}

\subsection{Comparison to Other Sparse Detection Methods}
Fig.\ \ref{fig:gmixMethComp} plots the squared-error performance
of the relaxed BP algorithm varying the measurement
ratio $\beta=n/m$ and holding $n=100$.
For each value of $\beta$, the
points labeled ``relaxed BP" show the median NSE after 20 iterations
of the relaxed BP algorithm.

Also plotted is the theoretical optimal MMSE performance
as predicted by the lower bound, Theorem \ref{thm:thetaLim}(c).
In this experiment, we observed only one
fixed point for the SE equations
for all the values of $\beta$.
So the lower bound on the MSE in Theorem \ref{thm:thetaLim}(c)
equals the theoretical asymptotic performance of relaxed BP
given in Theorem \ref{thm:thetaLim}(b).
We see that the relaxed BP algorithm at $n=100$
performs very close to the
asymptotic optimal performance for values of $\beta$ up to
approximately $2$.
For larger values of $\beta$, there is a small gap between
the performance of the relaxed BP algorithm
and the optimal performance.  The gap grows to 0.8 dB at $\beta=3$.
As discussed in Fig.\ \ref{fig:gmixDE}, this gap
decreases at higher values of $n$.

Fig.\ \ref{fig:gmixMethComp} also shows the performance
of two other simple algorithms.  The top curve is the
median NSE for optimal linear MMSE estimation,
and the curve labeled ``lasso" is the MSE from the lasso
algorithm of \cite{Tibshirani:96}.
The lasso method is based on an $\ell_1$-relaxation of the optimal
estimator and is widely-used for sparse estimation problems in compressed sensing.
In this experiment, the regularization weighting in the lasso estimator
was optimized as described in \cite{RanganFG:09arXiv}.

We see that the relaxed BP algorithm offers some gain over either of these methods.
Of course, with the interest in compressed sensing, there is now a plethora
of methods for estimating sparse vectors.
It is likely that other methods,
including possible modifications of lasso,
can obtain a similar performance as relaxed BP\@.
A complete comparison of relaxed BP
against these methods is beyond the scope of this work.
What is important is that relaxed BP provides a unified, systematic method for
a large class of problems, such that when applied to certain compressed sensing problems,
it gives near optimal performance.

\subsection{Estimation with Bounded Noise} \label{sec:bndSim}

To validate the relaxed BP method and analysis for non-AWGN output channels,
we next considered a bounded, uniform noise channel.
Specifically, we assumed that the output channel is given by
(\ref{eq:yzw}) where the components of the noise vector $\wbf$
are i.i.d. and uniformly distributed in an interval $[-\delta,\delta]$
for some $\delta > 0$.
Among other applications, this bounded noise model arises in
the study of subtractive dithered quantization
\cite{Gray:90b,LipshitzWV:92}, where the uncertainty interval
corresponds to a quantization region.

\begin{figure}
\begin{center}
  \includegraphics[width=3.5in]{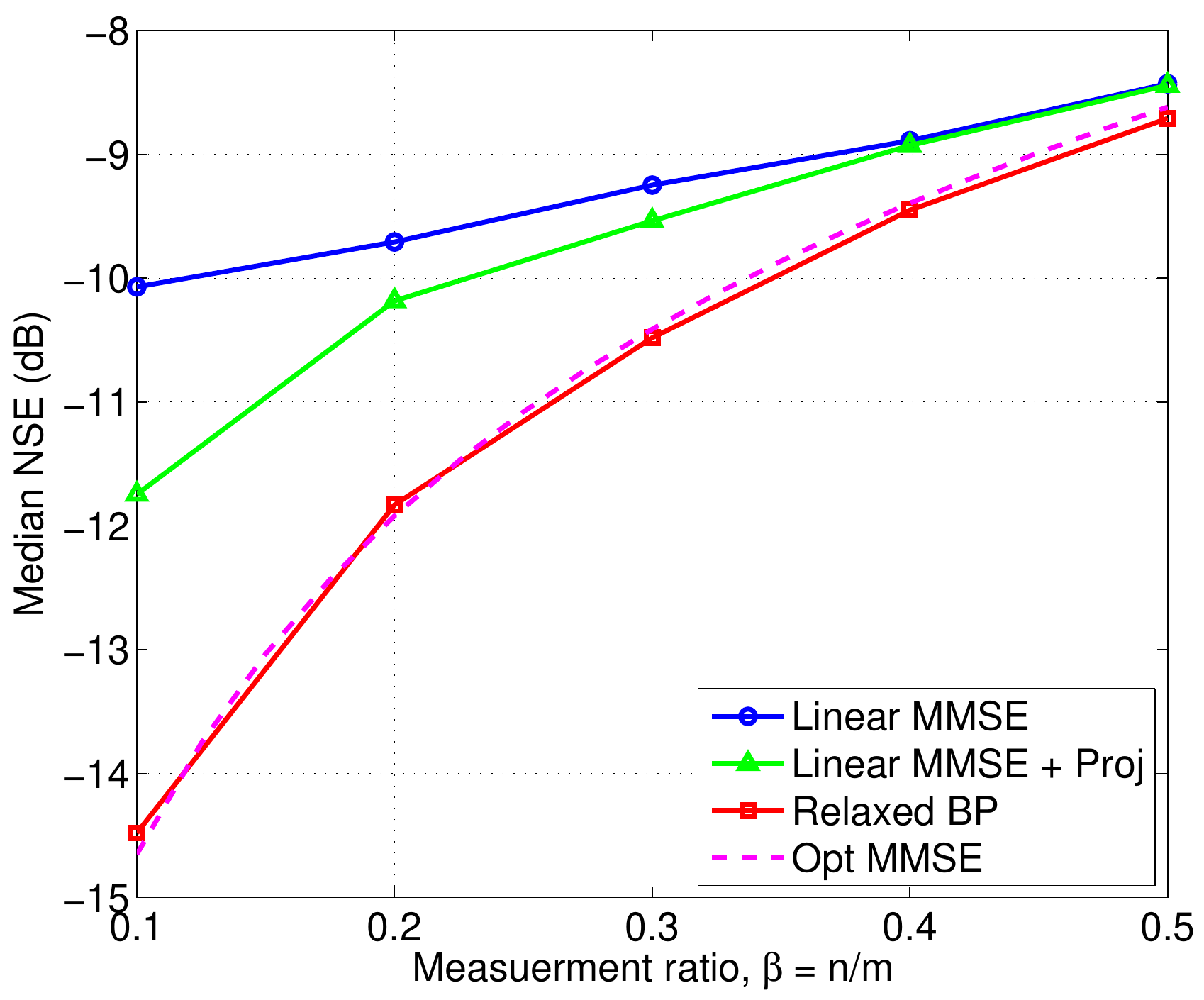}
\end{center}
\caption{Relaxed BP algorithm with a Gaussian prior and bounded noise
output channel.  The plot compares the simulated relaxed BP performance
against the predicted performance based on density evolution.
Also shown is the performance of the linear MMSE estimator
with and without projection to the consistent set.
}
\label{fig:lbpBndNoiseSim}
\end{figure}

Unfortunately, optimal MMSE estimation with bounded uniform
noise involves an integration
over an $n$-dimensional polytope, which is generally computationally intractable.
However, the relaxed BP algorithm can be readily applied to the relaxed BP problem
with bounded noise providing a simple, computationally-tractable
algorithm for this problem.

Fig.\ \ref{fig:lbpBndNoiseSim} shows a simulation of the relaxed BP algorithm
with a bounded uniform output noise channel.  The simulation used a
vector $\xbf$ with $n=50$ zero-mean i.i.d.\ Gaussian components.
Similar to the previous experiment, we again used a measurement matrix $\Phi$
with Gaussian i.i.d.\ components.  Also, bounded uniform noise in the interval
$[-\delta,\delta]$ results in a noise variance of $\mu_w = \delta^2/3$.
In this experiment, the noise level $\delta$ was adjusted such that $\SNR_0$ in
(\ref{eq:SNRdef}) was equal to 10 dB.  We varied the values of the
measurement ratio $\beta = n/m$, and for each value of $\beta$,
the points labeled ``Relaxed BP" in Fig.\ \ref{fig:lbpBndNoiseSim} plots
the median NSE over 1000 Monte Carlo trials of the relaxed BP algorithm,
using 20 iterations in each relaxed BP run.

As in the previous experiment, the SE equations have a unique fixed point,
and thus relaxed BP is theoretically asymptotically optimal with a
minimum variance predicted by the SE fixed point.
The curve labeled ``Opt MMSE" shows this theoretical
asymptotic minimum squared error.
We see that the median squared error of relaxed BP at $n=50$
matches the theoretical asymptotic performance well.

Fig.\ \ref{fig:lbpBndNoiseSim} also compares the relaxed BP method to
two other simple algorithms.
One is the linear MMSE estimator, which is equivalent to the MMSE estimator
assuming Gaussian noise.
The second estimator, shown in the curve labeled ``Linear MMSE + Proj," is the linear
MMSE estimate followed by a projection step.
A key observation of the work \cite{ThaoV:94,ThaoV:94b} is that
any estimate (including the linear MMSE estimate) can be improved by simply
projecting the estimate onto the set of vectors $\xbf$ \emph{consistent} with the
bounded noise.  An estimate $\xbfhat$ is consistent with the noise if
$\|\ybf-\Phi\xbfhat\|_\infty \leq \delta$.
The works \cite{ThaoV:94,ThaoV:94b} show that projecting to a consistent estimate
always reduces the squared-error and can offer significant gains at low values of $\beta$
(what is called high oversampling).
Similar results and algorithms have been reported elsewhere
\cite{ThaoV:96,Cvetkovic:99,RanganG:01}.
The figure shows that projecting the linear MMSE estimate does indeed offer
reductions in the squared error, especially for small $\beta$.
However, the relaxed BP algorithm, in comparison, is even better.

The reason that the relaxed BP algorithm shows a performance improvement
over the projected linear MMSE estimate is that
projecting the linear MMSE
estimate will generally result in a point only
on the boundary of the consistent set.
In contrast, the relaxed BP algorithm will attempt to find the centroid
of the consistent region, which will likely have a smaller error variance.

\section{Conclusions}
\label{sec:conclusions}
We have presented an extension to Guo and Wang's relaxed BP method in
 \cite{GuoW:06} to non-AWGN measurements.  The algorithm
applies to a large class of estimation problems involving
linear mixing and arbitrary separable input and output distributions.
Unlike standard BP, relaxed BP
is computationally tractable even for dense measurement matrices.
Our main result shows that, in the large sparse limit,
relaxed BP achieves the same
asymptotic behavior as standard BP as described in \cite{GuoW:07}.
In particular, when certain state evolution equations have unique fixed
points, relaxed BP is mean-square optimal.  Given the generality
of the algorithm, its computational simplicity and
provable performance guarantees, we believe that relaxed BP
can have wide ranging applications.  We have demonstrated the algorithm
in two well-known NP-hard problems:  compressed sensing and estimation
with bounded noise.

The main theoretical limitation of the work is that it applies
to large sparse random matrices, where the density of the measurement
matrix must grow at a much slower rate than the matrix dimension.
An interesting avenue of future work would be to see if the
dense matrix analysis of the AMP algorithm in \cite{DonohoMM:09arxiv} and
\cite{BayatiM:10arxiv} can be extended to relaxed BP\@.

\appendices

\section{Preliminary Convergence Results} \label{sec:modLimThm}

Before proving our main result, the next number of appendices
develop some preliminary results.
We begin in this appendix with some simple extensions to
the Law of Large Numbers and the Central Limit Theorem.
We omit the proofs as the results can be proven with minor modifications
to standard arguments using characteristic functions
\cite{GrimmettS:92}.

\begin{lemma}[Modified Law of Large Numbers] \label{lem:modLawLargeNum}
For each $n$ and $d$, let $x^d_{n,i} \in \R$,
$i=1,\ldots,d$ be a set of independent (though not necessarily identically distributed)
random variables satisfying
\[
    \lim_{d \arr \infty} \lim_{n \arr \infty} x^d_{n,i} = x \sim p_X(x),
\]
where the convergence is in distribution and $p_X(x)$ is
the distribution for the limiting random variable $x$
and $i=i(n,d) \in \{1,\ldots,d\}$ is any deterministic sequence.  Assume $x$
has bounded second moments.
Let $b^d_{n,i}$ be a set of non-negative deterministic constants such that
\[
    \lim_{d \arr \infty} \lim_{n \arr \infty}
        \frac{1}{d} \sum_{i=1}^d b^d_{n,i} = 1.
\]
Then, the limit
\[
    \lim_{d \arr \infty} \lim_{n \arr \infty}
    \frac{1}{d} \sum_{i=1}^d b^d_{n,i}x^d_{n,i} = \Exp(x)
\]
holds in distribution.
\end{lemma}

\begin{lemma}[Modified Central Limit Theorem] \label{lem:modCLT}
Let $x^d_{n,i}$ be as in Lemma \ref{lem:modLawLargeNum}
such that, for any deterministic sequence of indices
$i=i(n,d)\in\{1,\ldots,d\}$, we have the limit
\[
    \lim_{d \arr \infty} \lim_{n \arr \infty} \sqrt{d}|\Exp(x^d_{n,i})-\Exp(x)| = 0.
\]
Also, suppose that $a^d_{n,i}$ is a deterministic sequence of scalars such that
that
\beqan
    \lim_{d \arr \infty} \lim_{n \arr \infty}
        \frac{1}{d} \sum_{i=1}^d |a^d_{n,i}|^2 &=& 1, \\
    \lim_{d \arr \infty} \lim_{n \arr \infty} \frac{1}{d^{3/2}} \sum_{i=1}^d
        |a^d_{n,i}|^3 &=& 0.
\eeqan
Then,
\[
     \lim_{d \arr \infty} \lim_{n \arr \infty} \frac{1}{\sqrt{d}}
    \sum_{i=1}^d a^d_{n,i}(x^d_{n,i} - \Exp(x))
    =  {\cal N}(0, \var(x))
\]
where the limit is in distribution.
\end{lemma}

\section{Genie Algorithm} \label{sec:genieAlgo}
As stated earlier, part (c) of Theorem \ref{thm:thetaLim} is not new
and can be found in \cite{GuoW:06,GuoW:07}.
Their proof is restated here only for completeness.
Using similar arguments as \cite{RichardsonU:98},
their  proof considers a ``genie" or ``oracle-aided" algorithm
that has, as side information, knowledge of certain subsets of
the components $x_j$.  

The genie algorithm is defined as follows:
In step 1 of the relaxed BP algorithm
in Section \ref{sec:rbpalgo}, we simply replace the initialization
(\ref{eq:xmuInit}) at $t=1$ with
\begin{subequations} \label{eq:xmuInitGenie}
\beqa
    \xhat_{i \la j}(t) &=& \xhat_j(t) = x_j, \\
    \mu^x_{i \la j}(t) &=& \mu^x_j(t) = 0,
\eeqa
\end{subequations}
where $x_j$ is the true value of the component.
Otherwise, all the steps of the algorithm are the same
as the regular relaxed BP algorithm.

We will see that while the error in the regular BP algorithm
improves with each iteration, the genie algorithm starts
with zero error and then increases.  The performance
of the true optimal estimator is ``sandwiched" somewhere
between the genie and regular relaxed BP algorithms,
and thus consideration of the regular and relaxed algorithms
provide upper and lower bounds on the optimal performance.

\section{Proof of Lemma \ref{lem:mseDeriv}} \label{sec:mseDerivPf}
Fix $\mu > 0$ and for $r=0,1,2,\ldots$, define the functions
\beq \label{eq:Aidef}
    A_r(q) = \int x^r p_{X}(x) \phi(q-x \, ; \, \mu) \, dx.
\eeq
Then, the conditional mean $\FIn(q,\mu)$ and variance $\MseIn(q,\mu)$
are given by
\begin{subequations} \label{eq:FEA}
\beqa
    \FIn(q,\mu) &=& \frac{A_1(q)}{A_0(q)}, \\
    \MseIn(q,\mu) &=& \frac{A_2(q)}{A_0(q)} - \frac{A^2_2(q)}{A^2_0(q)}.
\eeqa
\end{subequations}
Now, taking the derivative of the Gaussian distribution
$\phi(\cdot \, ; \, \mu)$ in (\ref{eq:Gaussian}), it is easily verified that
\[
    \frac{\partial }{\partial q}\phi(q-x \, ; \, \mu) =
    \frac{x-q}{\mu}\phi(q-x \, ; \, \mu).
\]
Bringing this derivative inside the integral (\ref{eq:Aidef}) then
shows that
\beq \label{eq:Auderiv}
    \frac{\partial A_r(q)}{\partial q} = \frac{1}{\mu}\left(
        A_{r+1}(q) - qA_r(q)\right).
\eeq
Applying (\ref{eq:Auderiv}) to (\ref{eq:FEA}) we obtain
\beqan
    \lefteqn{\frac{\partial \FIn(q,\mu)}{\partial q} =
    \frac{\partial }{\partial q}\frac{A_1(q)}{A_0(q)} }\\
    &=&    \frac{(A_2(q)-qA_1(q))A_0(q) - (A_1(q)-qA_0(q))A_1(q)}{\mu A^2_0(q)}\\
    &=& \frac{A_2(q)}{\mu A_0(q)} - \frac{A^2_2(q)}{\mu  A^2_0(q)}
    = \frac{1}{\mu}\MseIn(q,\mu).
\eeqan

\section{Computation Subgraphs and Local Tree-like Properties}
\label{sec:localTree}

An essential assumption of the large sparse limit analysis is
that the Tanner graph is \emph{locally tree-like}.
To describe this property
more precisely, we review some standard definitions and results
that can be found in any description of BP such as
\cite{RichardsonU:09}.
Consider the Tanner graph $G$ for the linear mixing problem defined
in Section \ref{sec:bp}.
For $t=0,1,2,\ldots$, recursively define a sequence of \emph{computation subgraphs},
$G_j(t)$, $G_i(t)$, $G_{i \la j}(t)$ and $G_{i \ra j}(t)$, as follows:
\begin{enumerate}
\item \emph{Initialize:}  Set $t=1$, and for all $(i,j) \in E$ let
$G_{i \la j}(t)$ and $G_j(t)$ be the empty subgraphs.
\item \emph{Output update:}  For all $(i,j) \in E$, let
$G_{i \ra j}(t)$ be the subgraph containing, for all $r \in \NOut(i) \neq j$:
\begin{itemize}
\item[(a)] the edges $(i,r)$; and
\item[(b)] the subgraphs $G_{i \la r}(t)$.
\end{itemize}
Similarly, define the subgraph $G_i(t)$ to be the subgraph
using all $r \in \NOut(i)$.
\item \emph{Input update:}  For all $(i,j) \in E$, let
$G_{i \la j}(t+1)$ be the subgraph containing the node $x_j$ and
for all $\ell \in \NIn(j) \neq i$:
\begin{itemize}
\item[(a)] the edges $(\ell,j)$;
\item[(b)] the output nodes $y_\ell$; and
\item[(c)] the subgraphs $G_{\ell \ra j}(t)$.
\end{itemize}
Similarly, define the subgraph $G_j(t)$ to be the subgraph
using all $\ell \in \NIn(j)$.
Set $t=t+1$ and return to step 1.
\end{enumerate}

Now let $H_{i \ra j}(t)$ be the sigma algebra
generated by the components $y_\ell$ contained in the computation
subgraph $G_{i \ra j}(t)$.
Similarly, let $H_{i \la j}(t)$ be the sigma algebra
generated by the components $y_\ell$ contained in the computation
subgraph $G_{i \la j}(t)$.
To analyze the BP algorithm with the
``genie" initialization in Appendix \ref{sec:genieAlgo},
let $\HGenie_{i \ra j}(t)$ and $\HGenie_{i \la j}(t)$
be respectively the sigma algebras generated
by the entire vector $\ybf$ and the variable $x_r$
\emph{not} in the computation subgraphs $G_{i \ra j}(t)$
and $G_{i \la j}(t)$.
The following results are standard for BP
and can be proven using
arguments as in \cite{RichardsonU:01}.

\begin{lemma} \label{lem:sigAlg}  Consider
the sigma algebras $H_{i \la j}(t)$ and
$H_{i \ra j}(t)$ defined above.
\begin{itemize}
\item[(a)] In the standard BP algorithm in Section \ref{sec:bp},
the distributions $p^x_{i \la j}(t,x_j)$ are
$H_{i \la j}(t)$ measurable, and
the distribution $p^z_{i \ra j}(t,z_i)$
and likelihood function $p^u_{i \ra j}(t,u_j)$
are $H_{i \ra j}(t)$ measurable.
\item[(b)]
In the relaxed BP algorithm,
 $\xhat_{i \la j}(t)$ and $\qhat_{i \la j}(t)$
are $H_{i \la j}(t)$ measurable and
$\zhat_{i \ra j}(t)$ and $\uhat_{i \ra j}(t)$
are $H_{i \ra j}(t)$ measurable.
\end{itemize}
Similarly, under the oracle initialization described
in Appendix \ref{sec:genieAlgo},
the above statements hold with $H_{i \la j}(t)$ and $H_{i \ra j}(t)$
replaced by $\HGenie_{i \la j}(t)$ and $\HGenie_{i \ra j}(t)$.
\end{lemma}

\begin{lemma} \label{lem:localCond}
Consider the standard BP algorithm in Section \ref{sec:bp}
and the computation subgraphs defined above.
\begin{itemize}
\item[(a)]  If $G_{i \la j}(t)$ is a tree, then
$p^x_{i \la j}(t,x_j)$ is the conditional distribution of $x_j$ given
$H_{i \la j}(t)$.
\item[(b)] If $G_{i \ra j}(t)$ is a tree, then
$p^z_{i \ra j}(t,z_i)$ is the conditional distribution of
$z_{i \ra j}$ given $H_{i \ra j}(t)$.
\end{itemize}
Similarly, under the genie initialization,
the above statements hold with $H_{i \la j}(t)$ and $H_{i \ra j}(t)$
replaced by $\HGenie_{i \la j}(t)$ and $\HGenie_{i \ra j}(t)$.
\end{lemma}

\begin{lemma} \label{lem:localIndep}
Consider the relaxed BP algorithm in Section \ref{sec:rbp}
under the assumptions in Section \ref{sec:sparseLimAnal}.
\begin{itemize}
\item[(a)]  If $G_{i \la j}(t)$ is a tree, then
all the terms $(\zhat_{\ell \ra j}(t),\mu^z_{i \ra j}(t))$
are independent for different values $\ell \in \NIn(j) \neq i$.
\item[(b)] If $G_{i \ra j}(t)$ is a tree, then
the random vectors $\theta^x_{i \la r}(t)$
are independent for different values $r \in \NOut(i) \neq j$.
\end{itemize}
\end{lemma}

\section{Proof of Theorem \ref{thm:thetaLim}(a)} \label{sec:thetaLimAPf}

We prove this by induction.  It is clear that (\ref{eq:thetaxLim})
holds for $t=1$.  In part B below we will show that
if (\ref{eq:thetaxLim}) holds for some $t$, then so does
(\ref{eq:thetazLim}) and (\ref{eq:thetazTotLim}).
Then, in part C, we will show that if (\ref{eq:thetazLim})
holds for some $t$, then (\ref{eq:thetaxLim}) holds for $t+1$.
This will complete the induction argument.
Part A provides a preliminary calculation that we need in part C.

\subsection{Derivatives of the Score Function}

The following lemma characterizes the derivatives
$D_r(y,\zhat,\mu)$ in (\ref{eq:Drzdef}).
The result can also be found in \cite{GuoW:07},
but we sketch the proof here for completeness.
We will use this result below in the analysis of the output node update.

\begin{lemma} \label{lem:scoreDeriv}
Fix $\zhat$ and $\mu$ and consider random variables $y$
and $z$ generated by $z \sim {\cal N}(\zhat+u,\mu)$
and $y \sim p_{Y|Z}(y|z)$ for some $u \in \R$.
Consider the derivative of the score function in (\ref{eq:Drzdef}).
Then,
\begin{subequations}
\beqa
    \Exp\left[ D_1(y,\zhat,\mu) |u\right] &=&
    -u\Exp\left[D_2(y,\zhat,\mu)|u=0\right] \nonumber \\
    & & \; + \, O(u^2), \label{eq:DrExp} \\
    \var\left[ D_1(y,\zhat,\mu) |u\right] &=&
    \Exp\left[D_2(y,\zhat,\mu)|u=0\right] \nonumber \\
    & & \; + \, O(u^2).\label{eq:DrVar}
\eeqa
\end{subequations}
\end{lemma}
\begin{IEEEproof}
To simplify the notation, we will drop the dependence on $\zhat$ and $\mu$.
So, we will write, for example, $D_1(y)$ for $D_1(y,\zhat,\mu)$.
To prove (\ref{eq:DrExp}), we first note that
\beqa
    \lefteqn{\Exp\left[ D_1(y) |u\right] =
    \int p_{Y|U}(y|u)D_1(y) \, dy } \nonumber\\
    &=& \int p_{Y|U}(y|0)D_1(y) \, dy \nonumber \\
    & & \; + \, u\int \left. \frac{\partial}{\partial u}
    p_{Y|U}(y|u)\right|_{u=0}D_1(y) \, dy +O(u^2). \label{eq:expDoneFac}
\eeqa
Now, for the first term in (\ref{eq:expDoneFac}), note that
using the definition of $D_1(y)$ in (\ref{eq:Drzdef}), we have
\beqa
    \lefteqn{ \int p_{Y|U}(y|0)D_1(y) \, dy } \nonumber\\
    &=& -\int p_{Y|U}(y|0)\left. \frac{\partial}{\partial u}
    \log p_{Y|U}(y|u)\right|_{u=0} \, dy \nonumber\\
    &=& -\int \left. \frac{\partial}{\partial u}
    p_{Y|U}(y|u)\right|_{u=0} \, dy \nonumber\\
    &=& -\frac{\partial}{\partial u}\int
    \left. p_{Y|U}(y|u)\, dy \right|_{u=0} \nonumber\\
    &=& -\frac{\partial}{\partial u}(1) = 0. \label{eq:expDoneA}
\eeqa
Similarly, the second term in (\ref{eq:expDoneFac}) can be simplified by
evaluating the second-order derivative in the definition of $D_2(y)$ to obtain
\beqa
    \lefteqn{ \Exp\left[D_2(y)|u=0\right] } \nonumber\\
    &=& \int \frac{1}{p_{Y|U}(y|0)}\left(\frac{\partial}{\partial u}
    \left. p_{Y|U}(y|u)\right|_{u=0} \right)^2 \, dy \nonumber \\
    & & \; - \, \int \frac{\partial^2}{\partial u^2}
    \left. p_{Y|U}(y|u)\right|_{u=0} \, dy.   \label{eq:expDoneB}
\eeqa
Now,
\beqa
    \lefteqn{ \int \frac{\partial^2}{\partial u^2}
    \left. p_{Y|U}(y|u)\right|_{u=0} \, dy } \nonumber \\
    &=&  \frac{\partial^2}{\partial u^2}\int
    \left. p_{Y|U}(y|u)\right|_{u=0} \, dy
    =  \frac{\partial^2}{\partial u^2}(1) = 0.
     \label{eq:expDoneC}
\eeqa
Also,
\beqa
    \lefteqn{ \int \frac{1}{p_{Y|U}(y|0)}\left(\frac{\partial}{\partial u}
    \left. p_{Y|U}(y|u)\right|_{u=0} \right)^2 \, dy } \nonumber \\
     &=& \int \frac{\partial}{\partial u}
    \left. p_{Y|U}(y|u)\right|_{u=0} \frac{\partial}{\partial u}
    \left.  \log p_{Y|U}(y|u)\right|_{u=0} \, dy \nonumber \\
       &=& -\int \frac{\partial}{\partial u}
    \left. p_{Y|U}(y|u)\right|_{u=0} D_1(y) \, dy.  \label{eq:expDoneD}
\eeqa
Substituting (\ref{eq:expDoneC}) and (\ref{eq:expDoneD}) into (\ref{eq:expDoneB}),
we obtain that
\beq \label{eq:expDoneE}
    \Exp\left[D_2(y)|u=0\right] = -\int \frac{\partial}{\partial u}
    \left. p_{Y|U}(y|u)\right|_{u=0} D_1(y) \, dy.
\eeq
Then (\ref{eq:DrExp}) follows by substituting
(\ref{eq:expDoneA}) and (\ref{eq:expDoneE}) into (\ref{eq:expDoneFac}).
Equation (\ref{eq:DrVar}) is proved by similar manipulations.
\end{IEEEproof}

\subsection{Analysis of the Output Node Update}
\label{sec:thetaLimAOut}

Let $t \geq 1$ and suppose that (\ref{eq:thetaxLim}) holds for some $t$.
We will show that this induction hypothesis implies
(\ref{eq:thetazLim}) and (\ref{eq:thetazTotLim}).
We will just prove this implication for $t > 1$.
The proof for $t=1$ is similar.

Under the induction hypothesis (\ref{eq:thetaxLim}),
we first consider the
convergence of the terms $\mu^z_{i \ra j}(t)$.
From the factorization (\ref{eq:PhiFactor}) we have that
\beq \label{eq:PhiasOut}
    \Phi_{ij} = \frac{1}{\sqrt{d}}a_{ij}\sqrt{s_j}, \ \ \ \forall
        j \in \NOut(i).
\eeq
Using (\ref{eq:outLin}) and (\ref{eq:PhiasOut}), we have that
\beqa
    \mu^z_{i \ra j}(t) &=&
        \sum_{r \in \NOut(i) \neq j} |\Phi_{ir}|^2\mu^x_{i \la r} \nonumber \\
    &=&   \frac{1}{d}\sum_{r \in \NOut(i) \neq j} |a_{ir}|^2s_r\mu^x_{i \la r}(t).
        \label{eq:muijzSum}
\eeqa
By Lemma \ref{lem:localIndep}(b) and the assumption that $G_{i \ra j}(t)$
is a tree, the terms in the summation in (\ref{eq:muijzSum}) are independent.
Also, the induction hypothesis (\ref{eq:thetaxLim}) shows that
their asymptotic distribution is given by
\[
    \lim_{d \arr \infty} \lim_{n \arr \infty}
    s_r\mu^x_{i \la r}(t) =
        s\MseIn(q,\mu^q(t-1)/s),
\]
where the convergence is in distribution and
the random variables $s$ and $q$ are the terms in $\theta^x(t)$ in
(\ref{eq:thetax}).
For the regular relaxed BP algorithm $\mu^q(t-1) = \muHi^q(t-1)$
and, for the genie algorithm, $\mu^q(t-1) = \muLo^q(t-1)$.
In either case, the expectation of this limiting random variable is
\[
    \Exp\left[ s\MseIn(q,\mu^q(t-1)/s) \right] = \MseBarIn(\mu^q(t-1)),
\]
where $\MseBarIn(\mu^q(t-1))$ is defined in (\ref{eq:MseBarIn}).
Using (\ref{eq:Asql}), we can apply the Modified Law of Large Numbers
(Lemma \ref{lem:modLawLargeNum}),
to the sum in (\ref{eq:muijzSum}) to obtain
\beq
    \lim_{d \arr \infty} \lim_{n \arr \infty} \mu^z_{i \la j}(t)
        =    \beta \MseBarIn(\mu^q(t-1))
        = \mu^z(t), \label{eq:muzLim}
\eeq
where the convergence is in distribution and the last step follows
from the definition of $\mu^z(t)$ in (\ref{eq:muSE}).

We next consider the convergence of the variables $\zhat_{i \ra j}(t)$.
If we define $z_{i \ra j}$ as in (\ref{eq:zijBP}), then (\ref{eq:outLin})
and (\ref{eq:PhiasOut}) show that
\beqa
    \lefteqn{ z_{i \ra j} - \zhat_{i \ra j}(t) =
        \sum_{r \in \NOut(i)\neq j} \Phi_{ir}(x_r - \xhat_{i \la r}) } \nonumber \\
    &=&   \frac{1}{\sqrt{d}}\sum_{r \in \NOut(i)\neq j} a_{ir}\sqrt{s_r}(x_r - \xhat_{i \la r}).
    \label{eq:zijSum}
\eeqa
By Lemma \ref{lem:localIndep}(b) the
terms in the summation (\ref{eq:zijSum}) are independent.
Also assuming (\ref{eq:thetaxLim}) holds, the terms in the summation converge as
\[
    \lim_{d \arr \infty} \lim_{n \arr \infty} \sqrt{s_r}(x_r - \xhat_{i \la r})
    = \sqrt{s}(x-\FIn(q,\mu^q(t-1)/s)),
\]
where the convergence is in distribution and
the random variables $x$, $s$ and $q$ are the terms in $\theta^x(t)$ in (\ref{eq:thetax}).
The variances of the terms converge as
\beqa
    \lefteqn{ \lim_{d \arr \infty} \lim_{n \arr \infty}
    \Exp\left[s_r|x_r - \xhat_{i \la r}|^2\right] } \nonumber \\
    &=& \Exp\left[ s|x-\FIn(q,\mu^q(t-1)/s)|^2 \right] = \MseBarIn(\mu^q(t-1)). \nonumber
\eeqa
Using (\ref{eq:Asqr}), (\ref{eq:Acuber}) and (\ref{eq:muzLim}), we can
apply the Modified Central Limit Theorem (Lemma \ref{lem:modCLT})
to (\ref{eq:zijSum}) to obtain
\beq \label{eq:zhatLimA}
    \lim_{d \arr \infty} \lim_{n \arr \infty} z_{i \ra j} - \zhat_{i \ra j}(t)
    = {\cal N}(0,\mu^z(t)),
\eeq
where the convergence is in distribution.

Similarly one can show that
\begin{subequations} \label{eq:zhatLimB}
\beq
    \lim_{d \arr \infty} \lim_{n \arr \infty}  z_{i \ra j} = {\cal N}(0,\muInit^z)
\eeq
and
\beq
    \lim_{d \arr \infty} \lim_{n \arr \infty}  (z_{i \ra j} - \zhat_{i \ra j}(t))
        \zhat_{i \ra j}(t) = 0,
\eeq
\end{subequations}
where $\muInit^z$ is defined in (\ref{eq:muzInit}).
Equations (\ref{eq:zhatLimA}) and (\ref{eq:zhatLimB}) imply that
\beq \label{eq:zhatLimC}
    \lim_{d \arr \infty} \lim_{n \arr \infty} (z_{i \ra j},\zhat_{i \ra j}(t)) =
    (z,\zhat),
\eeq
where $z$ and $\zhat$ are the Gaussian random variables in (\ref{eq:zzhat})
with $\mu = \muavg_z(t)$.
Combining (\ref{eq:muzLim}) and (\ref{eq:zhatLimC}) proves (\ref{eq:thetazLim}).

This argument shows that if (\ref{eq:thetaxLim}) is true for some $t$,
then so is (\ref{eq:thetazLim}).  A similar argument shows that (\ref{eq:thetaxLim})
also implies (\ref{eq:thetazTotLim}), except we replace the summations over the
sets $r \in \NOut(i) \neq j$ with $r \in \NOut(i)$.

\subsection{Analysis of the Input Node Update}
For the next step in the induction proof, we want to prove that if (\ref{eq:thetazLim})
holds for some $t$, then (\ref{eq:thetaxLim}) holds for $t+1$.

Throughout this section, fix the input index $j$
and variables $s_j$ and $x_j$.
For each output index $\ell \in \NIn(j)$ and $u \in \R$,
define the Markov chain
\[
    \zhat^G_\ell \rightarrow z^G_\ell(u) \rightarrow y^G_\ell(u),
\]
where the random variables are distributed as
\begin{subequations}
\beqan
    \zhat^G_\ell &\sim& {\cal N}(0, \muInit^z-\mu^z(t)) \nonumber \\
    z^G_\ell(u) &\sim& {\cal N}(\zhat^G_\ell + u, \mu^z(t)) \nonumber \\
    y^G_\ell(u) &\sim& p_{Y|Z}(y|z^G_\ell(u)). \nonumber
\eeqan
\end{subequations}
Suppose the Markov chains are independent over different values of $\ell$.
We use the superscript ``G" here to indicate that the random
variables are Gaussian approximations to actual random variables in the
problems.
To be specific, first observe that
\beq \label{eq:PhiasIn}
    \Phi_{\ell j} = \frac{1}{\sqrt{d}}a_{\ell j}\sqrt{s_j}, \ \ \ \forall
        \ell \in \NOut(j).
\eeq
Combining (\ref{eq:PhiasIn}) with
the definition of $z_{i \ra j}$ in (\ref{eq:zijBP}) and
the fact that $\zbf = \Phi \xbf$, we have
\beq \label{eq:zellu}
    z_\ell = \sum_{r \in \NOut(\ell)} \Phi_{\ell j}x_j
        =  u_\ell + z_{\ell \ra j},
\eeq
where
\beq \label{eq:uellDef}
    u_\ell = \Phi_{\ell j}x_j = \frac{1}{\sqrt{d}}a_{\ell j}\sqrt{s_j}x_j.
\eeq
The induction hypothesis (\ref{eq:thetazLim}) and Lemma
\ref{lem:localIndep}(a) then show that
\beqa
    \lefteqn{
    \lim_{d \arr \infty} \lim_{n \arr \infty}
        (\zhat_{\ell \ra j}(t),z_{\ell},y_\ell,\mu^z_{\ell \ra j}(t))  }
        \nonumber \\
    &=& (\zhat^G_\ell, z_\ell^{G}(u_\ell), y_\ell^{G}(u_\ell), \mu^z(t)),
     \label{eq:zygLim}
\eeqa
where the convergence is in distribution.

With these definitions, we first consider the convergence of $\mu^q_{i \la j}(t)$.
Using (\ref{eq:outNLMu}), (\ref{eq:inLinMu}) and (\ref{eq:PhiasIn}), we have that
\beqa
    \lefteqn{ \frac{1}{\mu^q_{i \la j}(t)} =
    \sum_{\ell \in \NIn(j) \neq i} \frac{|\Phi_{\ell j}|^2}{\mu^u_{\ell \ra j}(t)} } \nonumber \\
    &=&   \frac{1}{d}\sum_{\ell \in \NIn(j) \neq i} |a_{\ell j}|^2s_j
        D_2(y_\ell, \zhat_{\ell \ra j}(t), \mu^z_{\ell \ra j}(t)). \label{eq:muqsum}
\eeqa
By Lemma \ref{lem:localIndep}(a), given $x_j$ and $s_j$,
all the terms in (\ref{eq:muqsum}) are independent.

Also, using (\ref{eq:zygLim}), we have the limit
\beqa
    \lefteqn{ \lim_{d \arr \infty} \lim_{n \arr \infty}
    D_2(y_\ell, \zhat_{\ell \ra j}(t), \mu^z_{\ell \ra j}(t)) } \nonumber \\
    &\stackrel{(a)}{=}& D_2(y^G_\ell(u_\ell), \zhat^G_\ell(u_\ell) \mu^z(t)) \nonumber \\
    &\stackrel{(b)}{=}& D_2(y^G_\ell(0), \zhat^G_\ell(0), \mu^z(t)) + O(|u_\ell|^3)
    \nonumber \\
    &\stackrel{(c)}{=}& D_2(y^G_\ell(0), \zhat^G_\ell(0), \mu^z(t))
    \label{eq:D2lim}
\eeqa
where the convergence in (a) is in distribution;
(b) follows from the assumption that $D_3(\cdot)$ is uniformly bounded
and (c) follows from the fact that (\ref{eq:uellDef}) shows that
$|u_\ell|^3 = O(d^{-3/2}) \arr 0$.
We can apply the Modified Law of Large Numbers
(Lemma \ref{lem:modLawLargeNum}) to the sum (\ref{eq:muqsum})
to obtain the limit
\beqa
    \lefteqn{ \lim_{d \arr \infty} \lim_{n \arr \infty} \frac{1}{\mu^q_{i \la j}(t)}
    } \nonumber \\
    &\stackrel{(a)}{=} &
    \lim_{d \arr \infty} \lim_{n \arr \infty}
    \frac{1}{d}\sum_{\ell \in \NIn(j) \neq i} |a_{\ell j}|^2s_j  \nonumber \\
    & & \qquad\qquad\qquad\qquad \times \, D_2(y^G_\ell(0), \zhat^G_\ell(0), \mu^z(t)) \nonumber \\
    &\stackrel{(b)}{=} &
    s_j\Exp\left[ D_2(y^G_\ell(0), \zhat^G_\ell(0), \mu^z(t)) \right] \nonumber \\
    &\stackrel{(c)}{=}& s_j\MseBarOut(\mu^z(t))
    \stackrel{(d)}{=} \frac{s_j}{\mu^q(t)},
    \label{eq:muqLim}
\eeqa
where the limit in (a) is in distribution and
follows from (\ref{eq:muqsum}) and (\ref{eq:D2lim});
(b) follows from (\ref{eq:Asql}) and the Modified Law of Large Numbers;
(c) follows from
the definition of $\MseBarOut(\cdot)$ in
(\ref{eq:MseBarOut}) and the fact that the expectation over $(y,z)$
in (\ref{eq:MseBarOut}) is identical to $(y_\ell^G(0), z_\ell^G(0))$;
and
(d) follows from the SE equation (\ref{eq:muSEq}).

We next turn to the distribution of $\qhat_{i \la j}(t)$.
Using (\ref{eq:outNL}), the update (\ref{eq:inLinq}) can be simplified to
\beqa
    \lefteqn{ \qhat_{i \la j}(t)
    \ = \ \mu^q_{i \la j}(t)
    \sum_{\ell \neq i}  \frac{\Phi_{\ell j}^*
        \uhat_{\ell \ra j}(t)}{\mu^u_{\ell \ra j}(t)} } \nonumber \\
    &=& -\mu^q_{i \la j}(t)\sum_{\ell \neq i} \Phi_{\ell j}^*
        D_1(y_\ell, \zhat_{\ell \ra j}(t),\mu^z_{\ell \ra j}(t)). \label{eq:qhatD}
\eeqa
So, using (\ref{eq:zygLim}) and (\ref{eq:muqLim}),
\beqa
    \lefteqn{ \lim_{d \arr \infty} \lim_{n \arr \infty}
    \qhat_{i \la j}(t) } \nonumber \\
    & = & \!\! -\frac{\mu^q(t)}{s_j}
    \lim_{d, n \arr \infty}
    \sum_{\ell \neq i} \Phi_{\ell j}^*
        D_1(y_\ell^G(u_\ell), \zhat_\ell^G(u_\ell),\mu^z(t)), \qquad
        \label{eq:qhatE}
\eeqa
where here and below we use the shorthand $\lim_{d,n}$ for $\lim_d \lim_n$.
Now define
\beqa
    e_\ell &=& D_1(y_\ell^G(u_\ell), \zhat_\ell^G(u_\ell),\mu^z(t))
    \nonumber \\
    & & \; + \, \Phi_{\ell j}x_j \Exp\left[ D_2(y_\ell^G(0), \zhat_\ell^G(0),\mu^z(t))
    \right], \label{eq:edef}
\eeqa
so we can rewrite (\ref{eq:qhatE}) as
\beqa
    \lefteqn{ \lim_{d \arr \infty} \lim_{n \arr \infty}
    \qhat_{i \la j}(t) }\nonumber \\
    &=& \!\! -\lim_{d, n \arr \infty}
    \sum_{\ell \neq i} \frac{1}{s_j} \Phi_{\ell j}^*e_\ell \nonumber \\
    & & \!\! - \lim_{d, n \arr \infty}
    \sum_{\ell \neq i}\frac{\mu^q(t)|\Phi_{\ell j}|^2x_j}{s_j} \Exp\left[
        D_2(y_\ell^G(0), \zhat_\ell^G(0),\mu^z(t)) \right] \nonumber \\
    &=& \!\! -\lim_{d, n \arr \infty}
    \frac{\mu^q(t)}{\sqrt{ds_j}}\sum_{\ell \neq i} a_{\ell j}^*e_\ell + x_j,
        \label{eq:qhatF}
\eeqa
where the last step follows from (\ref{eq:qhatE})
and (\ref{eq:PhiasOut}).

Now, applying Lemma \ref{lem:scoreDeriv} to $e_\ell$ in
(\ref{eq:edef}),
\beqa
    \Exp\left(e_\ell|u_\ell \right) &=& O(|u_\ell|^2) \nonumber \\
    \var\left(e_\ell|u_\ell \right) &=& \Exp\left[
        D_2(y_\ell^G(0), \zhat^G_\ell(0), \mu^z(t)) \right] + O(|u_\ell|^2) \nonumber\\
     &=&\frac{1}{\mu^q(t)} + O(|u_\ell|^2). \nonumber
\eeqa
From the definition of $u_\ell$ in (\ref{eq:uellDef}), the
$O(|u_\ell|^2)$ terms are $O(1/d)$ and thus can be ignored.
Applying the Modified Central Limit Theorem (Lemma \ref{lem:modCLT})
to the sum in (\ref{eq:qhatF}) along with (\ref{eq:Asql}) we obtain
\beqa
    \lim_{d \arr \infty} \lim_{n \arr \infty}
    \qhat_{i \la j}(t)
    &=& x_j + \frac{(\mu^q(t))^2}{s_j}{\cal N}\left(0,\frac{1}{\mu^q(t)}\right)
    \nonumber \\
    &=& {\cal N}\left(x_j,\frac{\mu^q(t)}{s_j}\right). \label{eq:qhatLimG}
\eeqa

Also since $x_j \sim p_X(x_j)$ and $s_j \sim p_S(s_j)$,
(\ref{eq:muqLim}) and (\ref{eq:qhatLimG}) together show that for any iteration $t$,
\beqa
    \lefteqn{ \lim_{d \arr \infty} \lim_{n \arr \infty} (x_j,s_j,
        \qhat_{i \la j}(t),\mu^q_{i \la j}(t)) } \nonumber \\
        &=& \left( x,s,{\cal N}\left(x, \frac{1}{s}\mu_q(t)\right), \frac{1}{s}\mu_q(t)\right),
        \label{eq:thetaqLim}
\eeqa
where the convergence is in distribution, $x \sim p_X(x)$ and $s \sim p_S(s)$.
Applying (\ref{eq:inNL}) to (\ref{eq:thetaqLim}) shows (\ref{eq:thetaxLim}).
Therefore, we have shown that if (\ref{eq:thetazLim}) holds for $t$,
then (\ref{eq:thetaxLim}) holds for $t+1$.
One can also show that if (\ref{eq:thetazLim}) holds for $t$,
then (\ref{eq:thetaxTotLim}) holds for $t+1$ using similar arguments
except we replace the summations over $\ell \in \NIn(j) \neq i$ with
$\ell \in \NIn(j)$.

\section{Proof of Theorem \ref{thm:thetaLim}(b) and (c)}
\label{sec:thetaLimBCPf}

Parts (b) and (c) of Theorem \ref{thm:thetaLim}
can be proven along the lines of Guo and Wang's analysis in
\cite{GuoW:06} and \cite{GuoW:07} using the concept
of an \emph{asymptotically sufficient statistic}
along with a standard \emph{sandwiching} argument.
Specifically, using their analysis, we will show that
the regular and ``genie" versions of the
relaxed BP algorithm provide sufficient statistics
for certain conditional distribution of the vectors $\xbf$ and $\zbf$.
The regular BP algorithm provides a sufficient statistic
relative to the sigma algebras $H_{i \la j}(t)$ and $H_{i \ra j}(t)$,
and the ``genie" relaxed BP algorithm in Appendix \ref{sec:genieAlgo}
provides a sufficient statistic relative to
$\HGenie_{i \la j}(t)$ and $\HGenie_{i \ra j}(t)$.
Moreover, the MSE relative to the sigma algebras is described by
the state evolution equations starting from the ``high" initial conditions
for the regular algorithm and ``low" initial condition for the
genie algorithm.
Since the sigma algebra generated by the actual observation vector $\ybf$
lies somewhere between these sigma algebras, $H$ and $\HGenie$,
the MSE  of the optimal estimator is ``sandwiched" between the two
solutions to the state evolution equations.

Since the arguments in this section follow very closely with Guo and
Wang's analysis in \cite{GuoW:06,GuoW:07}, we will just sketch the
proof.  Similar sandwiching arguments can be found in the early
analysis of LDPC codes in \cite{RichardsonU:98}.

We begin with the following definition.

\begin{definition} \label{def:asSuffStat}
Suppose
that $(x_n,q_n,H_n)$ is a sequence where, for every $n$,
$x_n$ and $q_n$ are random variables and $H_n$ is a sigma algebra.
We will say that $q_n$ is an \emph{asymptotically sufficient statistic}
for $x_n$ given $H_n$ with limiting distribution $(x_n,q_n) \arr (x,q)$ if:
\begin{itemize}
\item[(a)]  $q_n$ is $H_n$-measurable;
\item[(b)]  $(x_n,q_n) \arr (x,q)$ in distribution;
\item[(c)]  For any bounded continuous function $f(x)$,
\[
    \lim_{n \arr \infty} \left(\Exp( f(x_n) | H_n) - \Exp( f(x) |q=q_n)\right) = 0
\]
almost surely.
\end{itemize}
\end{definition}

The definition is a natural generalization of the concept of
a sufficient statistic.  Specifically, it says that the
conditional estimate $\Exp( f(x_n) | H_n)$ can be replaced
by $\Exp( f(x) | q=q_n)$ with asymptotically vanishing error.
That is, it is sufficient to use just $q_n$ instead of the entire
sigma algebra $H_n$ and use just the limiting distribution $(x,q)$
as opposed to the termwise distributions $(x_n,q_n)$.

Following along the lines of Guo and Wang
\cite{GuoW:06,GuoW:07}, we now prove the following.

\begin{theorem} \label{thm:asSuffStat}
For the relaxed BP algorithm:
\begin{itemize}
\item[(a)]  If $G_{i \ra j}(t)$ is a tree, then $\zhat_{i \ra j}(t)$
is an asymptotically sufficient statistic for $z_{i \ra j}$ given
$H_{i \ra j}(t)$ with the asymptotic distribution (\ref{eq:thetazLim}).
\item[(b)]  If $G_{i \la j}(t)$ is a tree, then $(\qhat_{i \la j}(t),s_j)$
is an asymptotically sufficient statistic for $x_j$ given
$H_{i \la j}(t)$ with the asymptotic distribution (\ref{eq:thetaqLim}).
\end{itemize}
The result also holds for the ``genie" algorithm in
Appendix~\ref{sec:genieAlgo} with $H_{i \ra j}(t)$ and $H_{i \la j}(t)$
replaced by $\HGenie_{i \ra j}(t)$ and $\HGenie_{i \la j}(t)$.
\end{theorem}

Similar to the proof of Theorem \ref{thm:thetaLim}(a),
we prove Theorem~\ref{thm:asSuffStat} by induction.
For the initial step in the induction, note that,
for the regular (non-Genie) algorithm,
$H_{i \ra j}(t)$ is empty and $\zhat_{i \ra j}(1)$ is the
prior on $z_{i \ra j}$.  For the genie algorithm,
$\HGenie_{i \ra j}(t)$ contains the entire vector $\xbf$ and
$\zhat_{i \ra j}(1) = z_{i \ra j}$.
Therefore, part (a) of Theorem \ref{thm:asSuffStat}
holds for $t=1$.
In part A below, we will show that if (a) holds for some $t$,
(b) holds for $t+1$.  In part B, we will show the reverse
implication that if (b) holds for some $t$ so does (a).
In part C, we apply Theorem \ref{thm:asSuffStat} to
prove Theorem \ref{thm:thetaLim}(b) and (c).

\subsection{Analysis of the Input Node Update}
Suppose that part (a) of Theorem \ref{thm:asSuffStat} holds
for some $t \geq 1$.  We will prove part (b) holds for $t+1$.
The asymptotic limit (\ref{eq:thetaqLim}) has already been proven.
We only need to show that $(\qhat_{i \la j}(t),s_j)$
is asymptotically sufficient to describe the conditional
distribution of $x_j$ given $H_{i\la j}(t+1)$.

To this end, suppose that $G_{i \la j}(t+1)$ is a tree.
By the construction of the computation subgraphs,
$G_{\ell \ra j}(t)$ must be a tree for every $\ell \in \NIn(j)$,
$\ell \neq j$.
Now define, for any $r \geq 1$,
the ``actual" derivatives of the likelihood
\beq \label{eq:Dract}
    D^{\ell \ra j}_{r,act}(t,y_\ell) = -\frac{\partial^r}{\partial u^r}
    \left. \log p^u_{\ell \ra j}(t,u) \right|_{u=0},
\eeq
where $p^u_{\ell \ra j}(t,u)$ is defined in (\ref{eq:puBP}).
Since $G_{\ell \ra j}(t)$ is a tree,
Lemma \ref{lem:localCond} shows that
$p^z_{\ell \ra j}(t,z_{\ell \ra j})$ in (\ref{eq:puBP}) is
the conditional distribution $z_{\ell \ra j}$ given $H_{\ell \ra j}(t)$.
Bringing the derivatives in (\ref{eq:Dract})
inside the expectation in (\ref{eq:puBP})
we can rewrite (\ref{eq:Dract}) as
\beqa
    \lefteqn{ D^{\ell \ra j}_{r,act}(t,y_\ell) } \nonumber \\
    &=& -\Exp\left[ \frac{\partial^r}{\partial u^r}
    \left. \log p_{Y|Z}(y_i|u+z_{\ell \ra j}) \right|_{u=0} \
    \mid H_{\ell \ra j}(t) \right], \nonumber \\
    \label{eq:DractExp}
\eeqa
where the expectation is over the conditional distribution
of $z_{\ell \ra j}$ given $H_{\ell \ra j}(t)$.
Also, using (\ref{eq:pzy}) and (\ref{eq:Drzdef}),
we can write
\beqa
    \lefteqn{ D_{r}(y,\zhat,\mu) } \nonumber \\
    &=& -\Exp\left[ \frac{\partial^r}{\partial u^r}
    \left. \log p_{Y|Z}(y_i|u+z) \right|_{u=0} \
    \mid \zhat,\mu \right], \label{eq:DrzExp}
\eeqa
where the expectation is over $z \sim {\cal N}(\zhat,\mu)$.

Now, the induction hypothesis, Theorem \ref{thm:asSuffStat}(a),
states that $\zhat_{\ell \ra j}(t)$ is
asymptotically sufficient for $z_{\ell \ra j}$ given $H_{\ell \ra j}(t)$
with the asymptotic distribution
\[
    \lim_{d \arr \infty} \lim_{n \arr \infty}
        (z_{\ell \ra j},\zhat_{\ell \ra j}(t)) = {\cal N}(0, P_z(\mu^z(t)),
\]
where $\mu^z(t) = \muHi^z(t)$ for the regular algorithm and
$\mu^z(t) = \muLo^z(t)$ for the ``genie algorithm".
Applying this property to (\ref{eq:DractExp}) to (\ref{eq:DrzExp}), we
obtain that
\beq \label{eq:Drdiff}
    \lim_{n \arr \infty} \lim_{d \arr \infty}
    D^{\ell \ra j}_{r,act}(t,y_\ell) -
    D_{r}(y_\ell,\zhat_{\ell \ra j}(t),\mu^z(t)) = 0,
\eeq
almost surely.

We can now rewrite $p^x_{i \la j}(t+1,x_j)$ in (\ref{eq:pxBP}) as
\beqan
    \lefteqn{ \lim_{d \arr \infty} \lim_{n \arr \infty}
    -\log p^x_{i \la j}(t+1,x_j) + \log p_X(x_j) + \mbox{const} } \nonumber \\
    &\stackrel{(a)}{=}& \lim_{d \arr \infty} \lim_{n \arr \infty}
    \sum_{\ell \in \NIn(j) \neq i} \log p^u_{\ell \ra j}(t,\Phi_{\ell j}x_j) \\
    &\stackrel{(b)}{=}& \lim_{d \arr \infty} \lim_{n \arr \infty}
     \sum_{\ell \in \NIn(j) \neq i}
    D^{\ell \ra j}_{r,act}(t,y_\ell)\Phi_{\ell j}x_j  \nonumber \\
    & & \; + \, \frac{1}{2} D^{\ell \ra j}_{r,act}(t,y_\ell)|\Phi_{\ell j}x_j|^2
    + O(|\Phi_{\ell j}x_j|^3) \nonumber \\
    &\stackrel{(c)}{=}& \lim_{d,n \arr \infty}
     \sum_{\ell \in \NIn(j) \neq i}
    D_{r}(y_\ell,\zhat_{\ell \ra j}(t),\mu^z(t))\Phi_{\ell j}x_j  \nonumber \\
    & & \; + \, \frac{1}{2} D_{r}(y_\ell,\zhat_{\ell \ra j}(t),\mu^z(t))|\Phi_{\ell j}x_j|^2
    \nonumber \\
    &\stackrel{(d)}{=}& \lim_{d,n \arr \infty}
     \frac{1}{2\mu^q_{i \la j}(t)}|x_j - \qhat_{i \la j}(t)|^2
    \nonumber \\
    &\stackrel{(e)}{=}& \lim_{d,n \arr \infty}
     \frac{s_j}{2\mu^q(t)}|x_j - \qhat_{i \la j}(t)|^2
\eeqan
where the constant is independent of $x_j$;
(a) follows from (\ref{eq:pxBP});
(b) is the Taylor's series expansion of $\log p^u_{\ell \ra j}(t,\Phi_{\ell j}x_j)$;
(c) follows from (\ref{eq:Drdiff}) and (\ref{eq:Acubel});
(d) follows from (\ref{eq:outNL}) and (\ref{eq:inLin}) and
(e) follows from (\ref{eq:muqLim}).
Here, with some abuse of notation, we have written $\lim A = \lim B$ in place
of $\lim (A-B) = 0$.
Using this same convention, the above equations show that
\beqa
    \lefteqn{ \lim_{d \arr \infty} \lim_{n \arr \infty}
    p^x_{i \la j}(t+1,x_j)  } \nonumber \\
    &=& \lim_{d \arr \infty} \lim_{n \arr \infty} \mbox{const } \nonumber \\
    & & \; \times \, p_X(x_j)\exp\left[
    \frac{1}{2\mu^q_{i \la j}(t)}|x_j - \qhat_{i \la j}(t)|^2 \right].
    \label{eq:pxLim}
\eeqa
From Lemma \ref{lem:localCond}, the left hand side of (\ref{eq:pxLim})
precisely the conditional distribution of $x_j$ given $H_{i \la j}(t+1)$
(or $\HGenie_{i \la j}(t)$).  Therefore, (\ref{eq:pxLim}) shows
that this conditional distribution is asymptotically only a function of
$\qhat_{i \la j}(t)$ and $s_j$, and therefore
$(\qhat_{i \la j}(t),s_j)$ is asymptotically sufficient
for $x_j$ given $H_{i \la j}(t+1)$.

\subsection{Analysis of the Output Update}

Continuing the induction argument,
we next show that if the part (b) of Theorem \ref{thm:asSuffStat} holds for
some $t$, then so does part (a).
We have already proven the asymptotic distribution (\ref{eq:thetazLim}).
So, we just need to show that the conditional
distribution of $z_{i \ra j}$ given $H_{i \ra j}(t)$
asymptotically depends only on $\zhat_{i \ra j}(t)$.

Now, from Lemma \ref{lem:localCond}, the conditional
distribution of $z_{i \ra j}$ given $H_{i \ra j}(t)$
is given by $p^z_{i \ra j}(t,z_{i \ra j})$ from the BP algorithm.
But this distribution is described by the summation (\ref{eq:zijBP}).
The analysis in Appendix \ref{sec:thetaLimAOut} shows that this
summation has an asymptotic Gaussian distribution
${\cal N}(\zhat_{i \ra j}(t), \mu^z(t))$.  So, $\zhat_{i \ra j}(t)$
is asymptotically sufficient to describe the distribution.

This completes the induction argument and proves Theorem~\ref{thm:asSuffStat}.

\subsection{MSE Relationships}
Using Theorem \ref{thm:asSuffStat}, we can now
prove parts (b) and (c) of Theorem \ref{thm:thetaLim}.
First observe that Theorem \ref{thm:asSuffStat} along with
the definition of an asymptotically sufficient statistic shows that
\beqa
    \lefteqn{
    \lim_{d \arr \infty} \lim_{n \arr \infty} \Exp(x_j|H_{i \la j}(t)) -
        \xhat_{i \la j}(t+1) }
    \nonumber \\
    &\stackrel{(a)}{=}&  \lim_{d \arr \infty} \lim_{n \arr \infty}
    \Exp( x | q=\qhat_{i \la j}(t), s=s_j)
        - \xhat_{i \la j}(t+1)\nonumber \\
    &\stackrel{(b)}{=}& \lim_{d \arr \infty} \lim_{n \arr \infty}
    \FIn(\qhat_{i\la j}(t), \mu^q(t)/s_j) - \xhat_{i \la j}(t+1)
    \nonumber \\
    &\stackrel{(c)}{=}& 0 \label{eq:ExpxLim}
\eeqa
where in (a) the expectation is with respect to $(x,q,s)$ distributed as
(\ref{eq:thetaqLim});
(b) follows from the definition of $\FIn(\cdot)$ in Section \ref{sec:scaEst};
and (c) follows from (\ref{eq:inNL}) and (\ref{eq:thetaqLim}).
The limit (\ref{eq:ExpxLim}) shows that the conditional variance is given by
\beqa
    \lefteqn{
    \lim_{d \arr \infty} \lim_{n \arr \infty}
        \Exp\left( \var(x_j|H_{i \la j}(t)) | s_j=s\right) }
    \nonumber \\
    &=&  \lim_{d \arr \infty} \lim_{n \arr \infty}
    \Exp\left(|x_j-\Exp(x_j|H_{i \la j}(t))|^2 | s_j=s\right) \nonumber \\
    &\stackrel{(a)}{=}&\lim_{d \arr \infty} \lim_{n \arr \infty}
        \Exp\left( |x_j-\xhat_{i \la j}(t)|^2 |s_j=s\right)  \nonumber \\
    &\stackrel{(b)}{=}& \Exp\left(|x-\FIn(q, \mu^q(t)/s)|^2 | s\right)\nonumber \\
    &\stackrel{(c)}{=}& \Exp\left(\MseIn(q,\mu^q(t)/s) | s\right)\nonumber \\
    &\stackrel{(d)}{=}& \MseBarIn(\mu^q(t),s)
    \stackrel{(e)}{=} \mu^x(t+1,s), \label{eq:mseLimxPf}
\eeqa
where (a) is due to the limit (\ref{eq:ExpxLim});
(b) is due to the limit (\ref{eq:thetaqLim});
(c) is the definition of $\MseIn(q,\mu)$;
(d) follows from (\ref{eq:MseBarIn}); and
(e) is from  (\ref{eq:muSEx}).
The limit (\ref{eq:mseLimxPf}) holds for the regular algorithm
with $\mu^x(t+1,s) = \muHi^x(t+1,s)$ and
for genie algorithm with $\mu^x(t+1,s) = \muLo^x(t+1,s)$.
With the regular (non-genie) algorithm, the limit (\ref{eq:mseLimxPf})
shows (\ref{eq:mseLimx}).  Also, for the genie algorithm, the sigma
algebra $\HGenie_{i \la j}(t)$ is contains the sigma
generated by just $\ybf$.
Therefore,
\[
    \var(x_j|\ybf) \geq \var(x_j|\HGenie_{i \la j}(t)).
\]
Combining this inequality with (\ref{eq:mseLimxPf}) shows
(\ref{eq:mseLimxLo}).

A similar argument can be used to show (\ref{eq:mseLimz}) and (\ref{eq:mseLimzLo}).
We have thus shown part (b) and (c) of Theorem \ref{thm:thetaLim}.

\section{Proof of Theorem \ref{thm:convIter}}
\label{sec:convIterPf}

The proof is based on a \emph{degradation} argument, which is used commonly
for convergence proofs of BP algorithms \cite{RichardsonU:09}.
Suppose that $X \arr Y \arr Z$ is a Markov chain.  Then, we say that $Z$ is \emph{degraded}
with respect to $Y$, since estimates of $X$ from $Z$ are strictly worse than those
from $Y$.  The following lemma states a standard property of degraded random variables.

\begin{lemma} \label{lem:varDeg}  Suppose that $X \arr Y \arr Z$ is a Markov chain.
Then,
\begin{itemize}
\item[(a)]  The conditional variance of $X$ satisfies
\[
    \var( X | Y ) \leq \var( X | Z).
\]
\item[(b)] Suppose the likelihood function of $X$ given $Y$ and
the likelihood of $X$ given $Z$ both have continuous third derivatives.
Then, for any $x$,
\[
    F(Z|X=x) \leq F(Y|X=x)
\]
where, for any random variables $X$ and $W$,  $F(W|X=x)$ is the Fisher information
\beq \label{eq:FisherDef}
    F(W|X=x) = -\Exp\left[ \left. \frac{\partial^2}{\partial x^2}\log p_{W|X}(w|x) \right|
        X = x \right]
\eeq
\end{itemize}
\end{lemma}
\begin{IEEEproof}  See, for example, \cite{LehmannC:03}.
\end{IEEEproof}

We will combine this lemma with the following simple iteration result to prove
the theorem.

\begin{lemma} \label{lem:monoConv}  Suppose $G(\mu)$ is a monotonically increasing,
continuous function with $0 \leq G(\mu) \leq \mu_{max}$ for all $\mu$ and
some $\mu_{max}$.
\begin{itemize}
\item[(a)] Consider the sequence $\mu(t+1) = G(\mu(t))$
initialized with $\mu(1) = \mu_{max}$.
Then
\[
    \lim_{t \arr \infty} \mu(t) = \mu,
\]
for some $\mu$ with $\mu = G(\mu)$.  Moreover, the limiting value $\mu$ is the
largest value satisfying $\mu = G(\mu)$ and $\mu \in [0,\mu_{max}]$.
\item[(b)]  Similarly, if the above sequence is initialized with $\mu(1) = 0$,
then $\mu(t) \arr \mu$ where $\mu$ is the smallest value satisfying
$\mu \in [0,\mu_{max}]$ and the fixed point equation $\mu = G(\mu)$.
\end{itemize}
\end{lemma}
\begin{IEEEproof}
We will just prove part (a) as part (b) is similar.
We first prove by, induction, that $\mu(t+1) \leq \mu(t)$ for all $t$.  For $t=1$,
since $\mu(1)$ is initialized to $\mu_{max}$ and $G(\mu) \leq \mu_{max}$ for all
$\mu$, we have that
\[
    \mu(2) = G(\mu(1)) \leq \mu_{max} = \mu(1).
\]
Now suppose that $\mu(t+1) \leq \mu(t)$ for some $t$.
Then, using the monotonicity of $G(\mu)$,
\[
    \mu(t+2) = G(\mu(t+1)) \leq G(\mu(t)) = \mu(t+1).
\]
So, by induction, $\mu(t)$ is a monotonically decreasing sequence.
Since it is bounded below by zero,
it must converge to some $\mu$.  By the continuity of $G(\mu)$, the limit point
must satisfy the fixed point equation $\mu = G(\mu)$.

It remains to show that the limiting value $\mu$ is the largest fixed point
of $G$ in the interval $[0,\mu_{max}]$.
To this end, let $\mu_1$ be any fixed point
$\mu_1 = G(\mu_1)$ with $0 \leq  \mu_1 \leq \mu_{max}$.  Then,
$\mu(1) = \mu_{max} \geq \mu_1$.  Also,
if $\mu(t) \geq \mu_1$, by the monotonicity of $G$,
\[
    \mu(t+1) = G(\mu(t)) \geq \mu_1.
\]
So, by the induction the entire sequence $\mu(t) \geq \mu_1$.  Taking the limits
as $t \arr \infty$, we have that $\mu \geq \mu_1$.
Hence $\mu \geq \mu_1$ for any other fixed point of $G$.
\end{IEEEproof}

We can now prove Theorem \ref{thm:convIter}.
Define the function
\beq \label{eq:GDE}
    G(\mu) = \beta \MseBarIn\left[ \MseBarOut(\mu)\right],
\eeq
so that we can rewrite the density evolution equation (\ref{eq:muSEOne}) as
\[
    \mu^z(t+1) = G(\mu^z(t)).
\]
We will now apply Lemma \ref{lem:monoConv} to
show that $\mu_z(t) \arr \mu$ to a fixed point $\mu = G(\mu)$.
We first upper bound $G(\mu)$.  For all $\mu > 0$,
\[
    \Exp\left[ \MseIn(q,\mu) \right]
    \stackrel{(a)}{=} \var(X|Q) \stackrel{(b)}{\leq} \var(X)
    \stackrel{(c)}{=} \muInit^x.
\]
where the expectation is over the random variable $q = x + v$,
$v \sim {\cal N}(0,\mu)$;
(a) follows from the definition of $\MseIn(q,\mu)$;
(b) is the fact that conditioning cannot increase the variance; and
(c) is from the definition of $\muInit^x$ in (\ref{eq:xmuInit}).
Therefore, the definition of $\MseBarIn(\mu)$ in (\ref{eq:MseBarIn}) implies that
\[
    \MseBarIn(\mu) = \Exp\left[ s \MseIn(q,\mu/s) \right]
     \leq \Exp(s) \muInit^x.
\]
As a result, $G(\mu)$ defined in (\ref{eq:GDE}) satisfies
\[
    G(\mu) \leq \beta \Exp(s)\muInit^x = \muInit^z,
\]
where $\muInit^z$ is defined in (\ref{eq:muzInit}).  So, we have that
$G(\mu) \leq \muInit^z$ for all $\mu$.  Also, the ``high"
sequence $\muHi^z(t)$
is initialized with $\muHi^z(1) = \muInit^z$
and the ``low" sequence with $\muLo^z(1) = 0$
So, we will apply Lemma \ref{lem:monoConv} with $\mu_{max} = \muInit^z$.

By the assumption of the theorem, $\MseBarIn(\mu)$ and $\MseBarOut(\mu)$
are continuous.  Therefore, so is $G(\mu)$.

Hence, to apply Lemma \ref{lem:monoConv},
it remains to show that $G(\mu)$ is monotonically increasing.
From (\ref{eq:GDE}), we need to show that
$\MseBarIn(\mu)$ and $\MseBarOut(\mu)$ are monotonically increasing.

We first consider $\MseBarIn(\mu)$.
Let $\mu_2 \geq \mu_1$ and define the random variables
\beqan
    q_1 &=& x + v_1, \ \ \ v_1 \sim {\cal N}(0,\mu_1/s), \\
    q_2 &=& q_1 + w, \ \ \ w \sim {\cal N}(0,(\mu_2-\mu_1)/s),
\eeqan
where $x \sim p_X(x)$, $s \sim p_S(s)$, and
$v_1$ and $v_2$ are independent.
We have that $x \arr q_1 \arr q_2$ is a Markov chain, so Lemma \ref{lem:varDeg}(a)
shows that, for all $s$,
\beq \label{eq:varXQ}
    \var(X|Q_1, S=s) \leq \var(X|Q_2, S=s).
\eeq
Also, $q_2$ is identically distributed to $q_2 = x + v_2$, $v_2 \sim {\cal N}(0,\mu_2/s)$
for some $v_2$ independent of $x$.
The definition of $\MseBarIn(\mu)$ shows that for $i=1,2$,
\beq \label{eq:mseXQ}
    \MseBarIn(\mu_i) = \Exp\left[ s \, \var(X|Q_i,S=s) \right].
\eeq
Combining (\ref{eq:varXQ}) and (\ref{eq:mseXQ}) shows that $\MseBarIn(\mu)$
is monotonically increasing in $\mu$.

The proof that $\MseBarOut(\mu)$ is monotonically increasing is similar.
Let $\mu_1$ and $\mu_2$ be variances such that
\[
    0 \leq \mu_1 \leq \mu_2 \leq \muInit^z.
\]
For $u \in \R$, define the random variables
\beqan
    \zhat_2 &\sim& {\cal N}(0,\muInit^z-\mu_2) \\
    \zhat_1 &\sim& \zhat_2 + {\cal N}(0,\mu_2-\mu_1) \\
    z &\sim& u + \zhat_1 + {\cal N}(0,\mu_1)
\eeqan
where all the Gaussian random variables are independent.
Also, conditional on $z$, let $y$ have the distribution $y \sim p_{Y|Z}(y|z)$.
It can be verified that
\[
    u \arr (\zhat_1,y) \arr (\zhat_2,y)
\]
is a Markov chain.  It follows from Lemma \ref{lem:varDeg}(b) that
\beq \label{eq:FmuIneq}
    F(\Zhat_1,Y|U=0) \geq F(\Zhat_2,Y|U=0).
\eeq
Also, the definitions of $z$, $\zhat_1$ and $\zhat_2$ above show that,
for $i=1,2$, when $u = 0$,
\[
    (z,\zhat_i) \sim {\cal N}(0,P_z(\mu_i)),
\]
where $P_z(\mu)$ is defined in (\ref{eq:Pzmu}).

Now, the Fisher information satisfies
\beqa
    \lefteqn{ F(\Zhat_i,Y|U=0) } \nonumber \\
    &\stackrel{(a)}{=}& -\Exp\left[ \left. \frac{\partial^2}{\partial u^2}
        \log p_{\Zhat_i,Y|U}(\zhat_i,y|u) \right| u=0 \right]. \nonumber \\
    &\stackrel{(b)}{=}& -\Exp\left[ \left. \frac{\partial^2}{\partial u^2}
        \log p_{Y|U,\Zhat_i}(y|u,\zhat_i) \right| u=0 \right]. \nonumber \\
    &\stackrel{(c)}{=}& \Exp\left[ D_2(y,\zhat_i,\mu_i) \right]. \nonumber \\
     &\stackrel{(d)}{=}& \frac{1}{\MseBarOut(\mu_i)}
     \label{eq:GOutFisher}
\eeqa
where (a) follows from the definition of the Fisher information in (\ref{eq:FisherDef});
(b) follows from the fact that
\beqan
   \lefteqn{ \log p_{\Zhat_i,Y|U}(\zhat_i,y|u) } \\
   &=& \log p_{Y|\Zhat_i,U}(y|u,\zhat_i) +
   \log p_{\Zhat_i|U}(\zhat_i|u)
\eeqan
and $\zhat_i$ is independent of $u$;
(c) is the definition of $D_r(y,\zhat,\mu)$ in (\ref{eq:Drzdef}); and
(d) follows from the definition of $\MseBarOut(\mu)$ in (\ref{eq:MseBarOut}).
Equations (\ref{eq:FmuIneq}) and (\ref{eq:MseBarOut}) together now
show that $\MseBarOut(\mu)$ is monotonically increasing in $\mu$.
Since $\MseBarIn(\mu)$ is also monotonically increasing in $\mu$, so is $G(\mu)$.

Lemma \ref{lem:monoConv} thus shows that $\muHi^z(t)$ converges to the largest
fixed point solution of the equation $\mu = G(\mu)$
and $\muLo^z(t)$ converges to the smallest fixed point.

\section*{Acknowledgments}
The author particularly thanks Dongning Guo and Chih-Chun Wang,
whose ideas are the basis of this work, and Vivek Goyal for their feedback.
The author also thanks Dror Baron,
Andrea Montanari, Phil Schniter, John Sun,
Lav Varshney and Martin Wainwright for their
careful reading of the manuscript and
Amin Shokrollahi and Martin Vetterli
for their generous support at the Summer Research Institute at EPFL, 2009,
where this research was initiated.

\bibliographystyle{IEEEtran}
\bibliography{bibl}

\end{document}